\documentclass[11pt]{article}
\usepackage{jheppub} 
 \usepackage[T1]{fontenc}
 \usepackage{booktabs,makecell}
\usepackage{colortbl, xcolor}
 \usepackage{xspace}
 \usepackage{dcolumn}
 \usepackage{hyperref}
 \usepackage[subrefformat=parens, position=top, skip=-15pt, margin=15pt, justification=justified, singlelinecheck=false] {subcaption}
 \usepackage{placeins}   
 \usepackage[group-digits=false]{siunitx}
 \usepackage{microtype}
 \usepackage{lmodern}
 \usepackage{hyperref}
\usepackage{multirow}
\usepackage{wasysym}

\pdfoutput=1

\allowdisplaybreaks[1]

\newcommand{\Pl}{\ell}

\newcommand{\fb}{{\ensuremath\unskip\,\text{fb}}\xspace}

\def\refeq#1{\mbox{(\ref{#1})}}
\def\reffi#1{\mbox{Fig.~\ref{#1}}}

\def\refta#1{\mbox{Table~\ref{#1}}}

\def\refse#1{\mbox{Section~\ref{#1}}}
\def\refses#1{\mbox{Sections~\ref{#1}}}

\def\citere#1{\mbox{Ref.~\cite{#1}}}
\def\citeres#1{\mbox{Refs.~\cite{#1}}}

\newcommand{\ri}{\mathrm i}

\newcommand{\ie}{{i.e.}\ }

\def\be{\begin{equation}}
\def\ee{\end{equation}}

\newcommand{\PH}{\ensuremath{\text{H}}\xspace}
\newcommand{\Pj}{\ensuremath{\text{j}}\xspace}
\newcommand{\PJ}{\ensuremath{\text{J}}\xspace}
\newcommand{\Pp}{\ensuremath{\text{p}}\xspace}

\newcommand{\Pb}{\ensuremath{\text{b}}\xspace}
\newcommand{\Pq}{\ensuremath{q}\xspace}
\newcommand{\Pt}{\ensuremath{\text{t}}\xspace}
\newcommand{\Pu}{\ensuremath{\text{u}}\xspace}
\newcommand{\Pd}{\ensuremath{\text{d}}\xspace}

\newcommand{\Pc}{\ensuremath{\text{c}}\xspace}
\newcommand{\Pg}{\ensuremath{\text{g}}\xspace}

\newcommand{\PW}{\ensuremath{\text{W}}\xspace}
\newcommand{\PZ}{\ensuremath{\text{Z}}}

\newcommand{\Mt}{\ensuremath{m_\Pt}\xspace}

\newcommand{\MW}{\ensuremath{M_\PW}\xspace}

\newcommand{\MZ}{\ensuremath{M_\PZ}\xspace}

\newcommand{\Gt}{\ensuremath{\Gamma_\Pt}\xspace}
\newcommand{\GH}{\ensuremath{\Gamma_\PH}\xspace}
\newcommand{\GZ}{\ensuremath{\Gamma_\PZ}\xspace}

\newcommand{\GW}{\ensuremath{\Gamma_\PW}\xspace}

\newcommand{\GeV}{\ensuremath{\,\text{GeV}}\xspace}
\newcommand{\TeV}{\ensuremath{\,\text{TeV}}\xspace}

\newcommand{\alphas}{\ensuremath{\alpha_\text{s}}\xspace}
\newcommand{\order}[1]{\ensuremath{\mathcal{O}{\left(#1\right)}}\xspace}

\newcommand{\GF}{\ensuremath{G_\mu}}

\newcommand{\newc}{\newcommand}
\newc{\bi}{\begin{itemize}}
\newc{\ei}{\end{itemize}}
\newc{\benu}{\begin{enumerate}}
\newc{\eenu}{\end{enumerate}}
\newc{\bc}{\begin{center}}
\newc{\ec}{\end{center}}
\newc{\bfig}{\begin{figure}}
\newc{\efig}{\end{figure}}
\newc{\qbar}{\bar{q}}
\newc{\go}{\tilde{g}}
\newc{\PB}{\textsc{Powheg-Box}}

\newcommand{\recola}{{\sc Recola}\xspace}

\newcommand{\mocanlo}{{\sc MoCaNLO}\xspace}

\newcommand{\bbmc}{{\sc BBMC}\xspace}
\newcommand{\collier}{{\sc Collier}\xspace}

\newcommand{\rT}{{\mathrm{T}}}
\newcolumntype{.}{D{.}{.}{-1}}
\newcolumntype{d}[1]{D{.}{.}{#1}}

\colorlet{tableoverheadcolor}{gray!37.5}
\colorlet{tableheadcolor}{gray!25}
\colorlet{tablerowcolor}{gray!12.5}

\newcommand{\gsim}
{\;\raisebox{-.3em}{$\stackrel{\displaystyle >}{\sim}$}\;}

\newlength{\width}
\newlength{\height}

\def\draftdate{\relax}
\def\mda{\relax}
\def\mua{\relax}
\def\mla{\relax}
\def\draft{
\def\thtystars{******************************}
\def\sixtystars{\thtystars\thtystars}
\typeout{}
\typeout{\sixtystars**}
\typeout{* Draft mode!
         For final version remove \protect\draft\space in source file *}
\typeout{\sixtystars**}
\typeout{}
\def\draftdate{\today}
\def\mua{\marginpar[\boldmath\hfil$\uparrow$]%
                   {\boldmath$\uparrow$\hfil}\color{black}%
                    \typeout{marginpar: $\uparrow$}\ignorespaces}
\def\mda{\color{red}\marginpar[\boldmath\hfil$\downarrow$]%
                   {\boldmath$\downarrow$\hfil}%
                    \typeout{marginpar: $\downarrow$}\ignorespaces}
\def\mla{\marginpar[\boldmath\hfil$\rightarrow$]%
                   {\boldmath$\leftarrow $\hfil}%
                    \typeout{marginpar: $\leftrightarrow$}\ignorespaces}
\def\Mua{\marginpar[\boldmath\hfil$\Uparrow$]%
                   {\boldmath$\Uparrow$\hfil}\color{black}%
                    \typeout{marginpar: $\uparrow$}\ignorespaces}
\def\Mda{\color{red}\marginpar[\boldmath\hfil$\Downarrow$]%
                   {\boldmath$\Downarrow$\hfil}%
                    \typeout{marginpar: $\downarrow$}\ignorespaces}
\def\Mla{\marginpar[\boldmath\hfil\textcolor{red}{$\Rightarrow$}]%
                   {\boldmath\textcolor{red}{$\Leftarrow $}\hfil}%
                    \typeout{marginpar: $\leftrightarrow$}\ignorespaces}
\overfullrule 5pt
\oddsidemargin 15mm
\marginparwidth 29mm
}

\newcommand{\pploprocr}{\Pp\Pp \to \ell^+ \nu_\ell\Pj\Pj\Pj\Pj}
\newcommand{\pploprocb}{\Pp\Pp \to \ell^+ \nu_\ell\Pj\Pj\PJ}
\newcommand{\qqloproc}{qq \to \ell^+ \nu_\ell qqqq}
\newcommand{\aaloproc}{\gamma\gamma\to \ell^+ \nu_\ell qqqq}
\newcommand{\qaloproc}{q\gamma \to \ell^+ \nu_\ell qqq\gamma}    

\title{\hfill ~\\[-58mm]
\phantom{h}\hfill\mbox{\small {}}
\\[1cm]
\vspace{13mm}   
NLO QCD and EW corrections to semileptonic vector-boson scattering 
at the LHC} 

\subheader{\today}

\author{Ansgar Denner$^1$,}
\author{Robert Franken$^1$,}
\author{Daniele Lombardi$^2$,}
\author{Santiago Lopez Portillo Chavez$^1$}

\affiliation{$^1$
        Institut f\"ur Theoretische Physik und Astrophysik, %
        Universit\"at W\"urzburg, \\ %
        Emil-Hilb-Weg 22,  %
        97074 W\"urzburg, %
        Germany%
}
\affiliation{$^2$ Dipartimento di Fisica, Universit\`a  di Torino and INFN, %
Sezione di Torino, \\
Via P. Giuria 1, 10125 Torino, Italy}

\emailAdd{ansgar.denner@uni-wuerzburg.de}
\emailAdd{robert.franken@uni-wuerzburg.de}
\emailAdd{daniele.lombardi@unito.it}
\emailAdd{santiago.lopez-portillo-chavez@uni-wuerzburg.de}

\abstract{Vector-boson scattering with semileptonic final
  states has recently been measured at the LHC, and future experiments are expected to
  further increase the precision of its measurement,
  calling for adequate theoretical predictions.
  In this work, we present a calculation of the NLO QCD and electroweak
  corrections to the process $\Pp\Pp \to \Pl^+ \nu_\Pl + 4\Pj$ in two
  different fiducial regions relevant for vector-boson scattering. In
  a fully off-shell calculation, we provide results for the leading
  electroweak contribution of $\order{\alpha^6}$ and the corresponding
  corrections of $\order{\alpha^7}$ and
  $\order{\alphas\alpha^6}$ for fiducial cross sections and a
  selection of differential distributions.  }

\begin{document} 

\maketitle

\section{Introduction}

Vector-boson scattering (VBS) constitutes a cornerstone of the rich research program
at the Large Hadron Collider (LHC). While all experimental data from
colliders seem to
confirm our current understanding of particle physics, based on
the Standard Model (SM), there are still parts of this theory that are poorly constrained
by measurements. In these areas, deviations from its predictions could arise,
hinting at physics 
beyond the SM, which,
despite of its successes, leaves many fundamental questions unanswered, such as the nature
of dark matter and dark energy as well as the origin of the
matter--antimatter asymmetry and the finite neutrino masses.
The electroweak (EW) sector of the SM includes parameters that have
been measured with unprecedented accuracy, and all its free parameters
are known. Nevertheless, in particular the Higgs sector of the
Lagrangian contains interaction terms for which no direct measurements
but only experimental upper bounds exist.

In this respect, despite its small cross sections and the large
background, VBS represents one of the most sensitive channels to probe
the EW symmetry-breaking (EWSB) mechanism and non-Abelian vector-boson
gauge interactions.  This is why VBS has attracted significant
attention from both the experimental and theoretical communities.
Its experimentally elusive nature is expected to become less challenging
once the full Run-3 dataset will be available, and even more so with
the upcoming high-luminosity stage of the LHC. In light of this,
theory predictions should also progress, providing increasingly
accurate results and better modelling for VBS, which remains a
non-trivial task.

Many VBS analyses have already been performed by both the ATLAS and CMS collaborations,
predominantly in the fully leptonic final state, \ie when the two scattering vector bosons
decay leptonically. In this way, in the last decade, essentially all scattering processes have been measured, namely
same-sign $\PW\PW$~\cite{
ATLAS:2019cbr,ATLAS:2023sua,CMS:2026gqm},
$\PW\PZ$~\cite{
CMS:2020gfh,ATLAS:2024ini,CMS:2026gqm},
$\PZ\PZ$~\cite{
CMS:2020fqz,ATLAS:2023dkz},
$\PW\gamma$~\cite{
CMS:2022yrl,ATLAS:2024bho}, and
$\PZ\gamma$~\cite{
CMS:2021gme,
ATLAS:2023fxh},
along with the more recent observation of $\PW^+\PW^-$ scattering~\cite{CMS:2022woe,ATLAS:2024ett}.
To further constrain the physics underlying VBS, new analyses addressing
polarised cross sections and polarisation fractions have begun to emerge. Such measurements
have been carried out for same-sign $\PW\PW$
scattering~\cite{CMS:2020etf,ATLAS:2025wuw}, as well as for
$\PW\PZ$~\cite{Aaboud:2019gxl,CMS:2021icx,ATLAS:2022oge,ATLAS:2024qbd}
and $\PZ\PZ$~\cite{ATLAS:2023zrv} scattering. Polarised
processes are known to be particularly sensitive to EWSB and potential unitarity violation at high
energy induced by new physics mechanisms. 

Given the strong interest in exploring the limits of the SM, VBS
analyses are moving towards higher precision using more data
and including more final states. In this respect,
VBS in the semileptonic final state, \ie where one vector boson decays
hadronically and one vector boson leptonically, plays a unique
role. The larger amount of background contaminating the signal region is partly compensated by
the larger branching ratio of a vector boson decaying into quarks as compared to leptons,
which makes the measurement of this final state feasible. Searches in this direction
have been pursued independently by CMS~\cite{CMS:2019qfk} and ATLAS~\cite{ATLAS:2019thr}.
The CMS collaboration provided evidence for semileptonic VBS already some years
ago~\cite{CMS:2021qzzv2}, while last year ATLAS reported its first observation of this
process~\cite{ATLAS:2025omi}.
To increase the sensitivity to regions of high transverse momentum of the vector bosons,
all semileptonic VBS studies consider two different categories. The partons arising
from the hadronically decaying vector boson are either reconstructed as separate jets or
clustered together into a single jet. The latter case covers events where the decay products of
the vector boson, owing to its large transverse momentum, are highly boosted and close in angle.

While measuring VBS is a delicate task, obtaining theory predictions for a process characterised
by a high final-state multiplicity and an intricate resonance
structure is challenging as well.
For a long time, the fully-leptonic decay mode was the only one that could be addressed.
Corresponding next-to-leading-order (NLO) QCD corrections, which are the simplest to compute,
had been known for quite some time, both for the EW production mode (which includes genuine VBS 
diagrams)~\cite{Jager:2006zc,Jager:2006cp,Bozzi:2007ur,Jager:2009xx,Denner:2012dz,Rauch:2016pai} and
for the QCD
background~\cite{Melia:2010bm,Melia:2011dw,Greiner:2012im,Campanario:2013qba,Campanario:2014ioa,Campanario:2014dpa}. Initial computations were performed in the \emph{VBS approximation},
which neglects $s$-channel diagrams along with the interference of $t$- and $u$-channel diagrams,
but these corrections are nowadays known in full generality~\cite{Denner:2012dz,Ballestrero:2018anz}.
Moreover, several results have been matched to parton-shower (PS)
codes~\cite{Jager:2011ms,Jager:2013mu,Jager:2013iza,Rauch:2016upa,Rauch:2016jxo,Jager:2018cyo,Jager:2024sij,Melia:2011tj}. At this perturbative order, VBS predictions have been obtained using very different
frameworks, like VBFNLO~\cite{Baglio:2014uba},
MadGraph5\_aMC@NLO~\cite{Stelzer:1994ta,Alwall:2014hca}, Sherpa~\cite{Sherpa:2019gpd},
and POWHEG-BOX~\cite{Nason:2004rx,Frixione:2007vw,Alioli:2010xd}.
A significant step forward was made when NLO EW corrections to fully leptonic final states
became available for all scattering
processes~\cite{Biedermann:2016yds,Denner:2019tmn,Denner:2020zit,Denner:2022pwc}.  For
some of them, even the full set of NLO contributions of $\mathcal{O}(\alphas^n\alpha^m)$ with $n+m=7$
was computed, namely for $\PW^+\PW^+$ scattering in~\citere{Biedermann:2017bss}
and for $\PZ\PZ$ in~\citere{Denner:2021hsa}. These results, which account for all
off-shell and interference effects, were obtained using the Monte Carlo integrator
\mocanlo~\cite{Denner:2026phn,denner_2026_19829093}.
As opposed to QCD corrections, the matching of NLO EW corrections for VBS to PS is still a relatively unexplored research area, and
only in~\citere{Chiesa:2019ulk} NLO EW predictions for same-sign $\PW$
scattering were matched to a QED PS and interfaced to a QCD PS.
Moreover, results for polarised cross sections also started to appear, initially just for
leading-order (LO) contributions,
in the {\sc Phantom} and {\sc MadGraph5\_aMC@NLO} event
generators~\cite{Ballestrero:2020qgv,Ballestrero:2017bxn,Ballestrero:2019qoy,BuarqueFranzosi:2019boy}.
Very recently, NLO QCD and EW corrections for polarised  $\PW^+\PW^+$ scattering ~\cite{Denner:2024tlu},
and NLO EW corrections for polarised $\PW\PZ$ scattering ~\cite{Denner:2025xdz} became available with \mocanlo.

While the amount of predictions for VBS in the fully leptonic final state is
remarkable, the literature for the semileptonic final state remains sparse.
This gap largely stems from the considerable efforts required to handle the calculation in full generality.
Since the nature of the hadronically decaying vector boson cannot be directly identified in experiments
through its decay products, a proper computation must account for multiple VBS processes simultaneously.
This results in a large number of partonic channels opening up already
at LO, and even more at NLO.
In \citeres{Jager:2013mu,Jager:2024sij}, the QCD corrections to the EW LO, \ie
$\mathcal{O}(\alphas \alpha^6)$, were computed in the VBS approximation separately for
$\Pp \Pp \to \PW^+ \PW^- \Pj\Pj$ and $\Pp \Pp \to \PW^+ \PZ \Pj\Pj$
in the semileptonic final states and matched to QCD parton showers.
Apart from this, only two works have presented results for semileptonic VBS without approximations,
where only LO contributions were addressed. In~\citere{Ballestrero:2008gf}, three out of five
LO contributions to $\Pp\Pp \to \ell^+ \nu_{\ell} + 4\Pj$, \ie
$\mathcal{O}(\alpha^6)$, $\mathcal{O}(\alphas^2\alpha^4)$ and $\mathcal{O}(\alphas^4\alpha^2)$,
were obtained using the {\sc Phantom} generator. In~\citere{Denner:2024xul}, the three LO
contributions $\mathcal{O}(\alpha^6)$, $\mathcal{O}(\alphas^2\alpha^4)$, and $\mathcal{O}(\alphas\alpha^5)$
to the same VBS process were computed using the \mocanlo{} code. In the latter work, the
complete off-shell calculation was compared to the double--pole-approximated
one defined according to~\citeres{Denner:2005fg,Denner:2000bj},
which allowed to estimate the importance of off-shell effects.
Given the growing interest in this signal within the experimental community, it is becoming increasingly
relevant to understand to what extent theory predictions can be improved by including higher-order corrections
in the perturbative expansion of the cross section or by matching exact VBS results to PS generators.

In this work, we take a significant step forward in accuracy by assessing for the first time
full off-shell NLO corrections in both EW and QCD to the $\mathcal{O}(\alpha^6)$
for the process $\Pp\Pp \to \ell^+ \nu_{\ell}  + 4\Pj$, namely by
computing the $\mathcal{O}(\alpha^7)$ and $\mathcal{O}(\alphas\alpha^6)$ contributions.
The calculation is  performed with the publicly available \mocanlo{}
integrator~\cite{Denner:2026phn,denner_2026_19829093} and validated against the independent in-house code \bbmc.
Results are presented in the two different fiducial regions introduced
in \citere{Denner:2024xul}, one where the decay products of the
hadronically-decaying vector boson are resolved and one where this
vector boson is reconstructed as a fat jet.

This manuscript is organised as follows. In \refse{se:description}, we present the general
structure of the calculation and we specify the partonic contributions included in our results.
Details on the evaluation of NLO EW and QCD corrections are discussed in
\refses{se:nloew} and \ref{se:nloqcd}, respectively. Neglected
contributions are commented on in \refse{se:neglected}. In \refse{se:validation}, we explain how we validated
our \mocanlo{} calculation using \bbmc. The input parameters and the definition of the
fiducial volume are specified in \refses{se:inputs} and \ref{se:cuts}, while in
\refse{se:results} we present results for the integrated and differential cross sections.
We conclude in \refse{conclusions}.

\section{Description of the calculation}
\label{se:description}

In this article we investigate semileptonic VBS with a leptonically
decaying $\PW^+$ boson. At the LHC, this class of processes receives
contributions from partonic processes of the type
\begin{equation}\label{eq:lo_subproc}
        \qqloproc \,,
\end{equation}
at lowest perturbative order, where $q$ stands for an arbitrary quark
or antiquark, $\ell^+$ is a positively charged lepton and $\nu_{\ell}$ the corresponding neutrino.  
In the phase-space region where the VBS contribution is dominant, which is
characterised by two jets with high invariant mass and rapidity separation,
the partonic processes in \refeq{eq:lo_subproc}
give rise to the physical processes
\begin{align}
        &\pploprocr \quad \text{ and} \label{eq:phys_proc_res} \\
        &\pploprocb \label{eq:phys_proc_boo}\,,
\end{align}
where $\Pj$ denotes
a jet with ordinary recombination radius $R$ (typically $R=0.4$)
and $\PJ$ a fat jet with a larger $R$ and high invariant mass.
In this article, we investigate both of these physical processes by employing two different sets of selection cuts, which we detail in \refse{se:cuts}.  

While doubly-photon-induced partonic processes of the type
\begin{equation}\label{eq:lo_subproc_aa}
        \aaloproc
\end{equation}
enter the physical processes in Eqs.~(\ref{eq:phys_proc_res}) and (\ref{eq:phys_proc_boo}) at the same perturbative order as those in \refeq{eq:lo_subproc}, we neglect them in this calculation owing to the small size of their contribution (see \refse{se:ph-induced}).
For the same reason, we also omit results for the single-photon-induced channels
\begin{equation}\label{eq:lo_subproc_aq}
        \qaloproc
\end{equation}
which, due to the insufficient number of final-state partons,
can only contribute to \refeq{eq:phys_proc_boo} with events where the final-state photon
recombines with one of the partons into a fat jet $\PJ$ (see \refse{se:ph-induced}).

\begin{figure}
\centering
\begin{subfigure}{0.3\textwidth}
\captionsetup{skip=0pt}
\caption{}
\centering
\includegraphics[page=6,width=1.\linewidth]{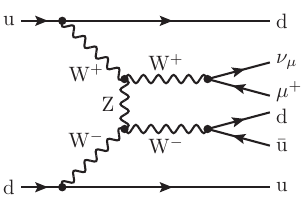}
\end{subfigure}\quad
\begin{subfigure}{0.3\textwidth}
\captionsetup{skip=0pt}
\caption{}
\centering
\includegraphics[page=7,width=1.\linewidth]{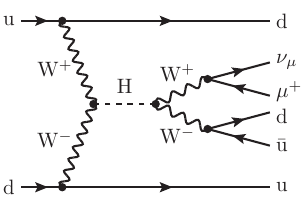}
\end{subfigure}\quad
\begin{subfigure}{0.3\textwidth}
\captionsetup{skip=0pt}
\caption{}
\centering
\includegraphics[page=8,width=1.\linewidth]{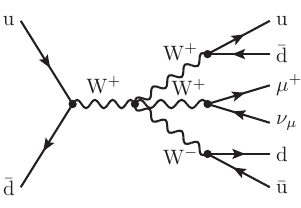}
\end{subfigure}
\caption{Examples of LO diagrams of $\order{g^6}$
  for partonic processes of type $\Pq\Pq/\Pq\bar\Pq \to \Pl^+\nu_\Pl 4\Pq$,
  explicitly $\Pu\Pd \to \mu^+ \nu_\mu \Pu\bar\Pu \Pd\Pd$ or
  $\Pu\bar\Pd \to \mu^+ \nu_\mu \Pu\bar\Pu \Pd\bar\Pd$. Diagram (a) represents
  a typical VBS topology, diagram (b) involves a resonant Higgs boson,
  and diagram (c) contributes to triple vector-boson production.}
\label{fig:lo}
\end{figure}

At LO, the physical processes \refeq{eq:phys_proc_res} and
\refeq{eq:phys_proc_boo} receive contributions of $\order{\alpha^6}$, $\order{\alphas\alpha^5}$, $\order{\alphas^2\alpha^4}$, $\order{\alphas^3\alpha^3}$, and $\order{\alphas^4\alpha^2}$. 
Genuine VBS topologies, where two quark lines interact via the emission of two EW bosons, 
are exclusively contained in the
$\order{\alpha^6}$. Some sample Feynman diagrams relevant at this perturbative order are shown in \reffi{fig:lo}.
In \citere{Denner:2024xul}, the first three orders were investigated in detail, including a discussion of their resonance structure and the irreducible background. In the present work, we build on this investigation and calculate, besides the LO contribution of $\order{\alpha^6}$, the NLO corrections of $\order{\alpha^7}$ and $\order{\alphas\alpha^6}$. 
The contribution of $\order{\alpha^7}$ contains only EW corrections to
the $\order{\alpha^6}$, whereas the one of $\order{\alphas\alpha^6}$ comprises both QCD corrections to the $\order{\alpha^6}$  and 
EW corrections to the $\order{\alphas\alpha^5}$. 
The calculation presented in \citere{Denner:2024xul} revealed that the contribution  of the $\order{\alphas\alpha^5}$
to the total cross section amounts to only 3\% of the one of $\order{\alpha^6}$.
Therefore, we expect the $\order{\alphas\alpha^6}$ contributions to be
strongly dominated by the QCD corrections to the EW LO process, and
often refer to them as such for short.

\subsection{EW corrections: \texorpdfstring{$\order{\alpha^7}$}{O(alpha7)}}
\label{se:nloew}

\begin{table}
\centering
\begin{tabular}{l|cc}
\hline
\rowcolor{blue!30}
\multicolumn{1}{c}{class}
&\multicolumn{1}{c}{partonic processes} \\
\hline
                    $\Pq\Pq                \to \Pl^+\nu_\ell \,  4 \Pq \gamma$ & 202 \\ 
\rowcolor{black!10} $\gamma\gamma \to \Pl^+\nu_\ell  \, 4 \Pq \gamma$   & 8   \\ 
\rowcolor{black!10} $\Pq\gamma         \to \Pl^+\nu_\ell  \, 5 \Pq $                & 96  \\ 
\rowcolor{black!20} $\Pq\gamma         \to  \Pl^+\nu_\ell \, 3 \Pq 2\gamma$ & 28  \\ 
\hline
\end{tabular}
\caption{Classes and number of partonic processes contributing to the
  real EW corrections of $\order{\alpha^7}$, where $\Pq$ stands for
  both quarks and antiquarks. Note that we count
  reflected initial states ($\Pq\Pq'/\Pq'\Pq$ or
  $\Pq\gamma/\gamma\Pq$) as one channel, even if these need separate
  calculations. Rows that are greyed out indicate contributions
  neglected in our final results. Dark grey contributions are
  discussed in \refse{se:misidentification}, light grey ones in
  \refse{se:ph-induced}.} 
\label{tab:processes_EW}
\end{table}

Four classes of partonic processes contribute to the real corrections
at $\order{\alpha^7}$, which are summarised in \refta{tab:processes_EW}. The first and second
classes correspond to real-photon emission off the LO processes in Eqs.~(\ref{eq:lo_subproc}) and (\ref{eq:lo_subproc_aa}), respectively. 
The third class is obtained from the first one by crossing the
final-state photon into the initial state. Finally, processes in the fourth
class result from real-photon emission off the LO process in \refeq{eq:lo_subproc_aq}.
Partonic channels in this last class have two photons and only three quarks in the final state. Since
we do not treat photons as jets, they
cannot contribute to the physical process in \refeq{eq:phys_proc_res}.
However, analogously to their LO channel, they may contribute to the physical process
in \refeq{eq:phys_proc_boo} if a photon and a quark recombine into a massive jet $\PJ$. 

The first class of partonic processes yields the largest contribution because of the strong dominance of the parton distribution functions (PDFs) of the quarks over the photon PDF. The results presented in \refse{se:results} include only this class of real corrections. As confirmed by the estimate in \refse{se:neglected},
the other three classes are negligible and can be safely dropped from the calculation. 

Consistently with the approximation used for the real contribution,
only virtual corrections of $\order{\alpha^7}$ to the $\Pq\Pq$-induced
channel are included, while the ones to the $\gamma\gamma$- and $\gamma q$-induced channels
are omitted. Including all contributions for a given initial state consistently ensures the infrared safety of our results.

\subsection{QCD corrections:
  \texorpdfstring{$\order{\alphas\alpha^6}$}{O(alpha\_s alpha6}}
\label{se:nloqcd}             
\begin{table}
\centering
\begin{tabular}{l|c}
\hline
\rowcolor{blue!30}
\multicolumn{1}{c}{class} &
\multicolumn{1}{c}{partonic processes} \\
\hline
                              $\Pq\Pq                  \to \Pl^+\nu_\ell \,  4 \Pq \Pg$                & 202  \\
\rowcolor{black!10}$\gamma \gamma \to \Pl^+\nu_\ell \,  4\Pq  \Pg$                & 8  \\
                              $\Pq\Pg                  \to \Pl^+\nu_\ell \,  5 \Pq$                      & 96  \\
\rowcolor{black!10}$\Pq\gamma          \to \Pl^+\nu_\ell \,  3\Pq \Pg \gamma$   & 28 \\
\rowcolor{black!10}$\Pg\gamma          \to \Pl^+\nu_\ell \,  4\Pq \gamma$          & 8 \\
\rowcolor{black!20}$\Pq\Pg                  \to\Pl^+\nu_\ell \,  3\Pq 2\gamma$        & 28 \\
\rowcolor{black!20}$\Pq\Pq                  \to \Pl^+\nu_\ell \, 2\Pq \Pg 2 \gamma$ & 78 \\  \hline
                              $\Pq\Pq                   \to\Pl^+\nu_\ell \,  4 \Pq \gamma$         & 202 \\ 
\rowcolor{black!10}$\gamma \gamma  \to \Pl^+\nu_\ell \, 4\Pq  \gamma$
& 4 \\  
\rowcolor{black!10}$\Pq\gamma           \to \Pl^+\nu_\ell \, 5\Pq $                       & 96 \\
\rowcolor{black!20}$\Pq\gamma           \to \Pl^+\nu_\ell \, 3\Pq
2\gamma$        & 12 \\  
\hline
\end{tabular}
\caption{Classes and numbers of partonic processes contributing to the
  real QCD corrections of $\order{\alphas\alpha^6}$. Note that we count
  reflected initial states as one channel (see \refta{tab:processes_EW}) and that some
  partonic processes that vanish owing to $\mathrm{SU}(3)$ colour
  algebra are already excluded in the counting. Rows that are greyed
  out indicate contributions neglected in the final results. Dark grey contributions are
  discussed in \refse{se:misidentification}, light grey ones in
  \refse{se:ph-induced}.} 
\label{tab:processes_QCD}
\end{table}

Owing to the possible presence of an external gluon, 
partonic processes for the real corrections at
$\order{\alphas\alpha^6}$ show a larger variety than in the case of EW
corrections. They amount to eleven different classes, which are listed in
\refta{tab:processes_QCD}. The first seven classes, which involve an
external gluon, can be identified as pure QCD corrections to the $\order{\alpha^6}$, since their
contribution consists of squared matrix elements. The remaining four
classes, which have no external gluons, must contain an internal gluon
to contribute to this order and thus result from interferences between diagrams of $\order{g_\mathrm s^2 g^5}$ and of $\order{g^7}$,
\ie EW corrections to the $\order{\alphas\alpha^5}$.

The first and second classes in each group are 
obtained from a LO process by adding a gluon or photon. The others are constructed
by crossing initial- and final-state partons. Like in the case of the EW corrections,
there are processes with two final-state photons contributing only
to the physical process in \refeq{eq:phys_proc_boo}.

Out of the eleven different classes, only three give rise to a sizeable
contribution or are otherwise relevant: $\Pq\Pq \to \Pl^+\nu_\ell 4\Pq\Pg$, $\Pq\Pg \to \Pl^+\nu_\ell5\Pq$ and $\Pq\Pq \to \Pl^+\nu_\ell4\Pq\gamma$. The first two, pure QCD corrections, are expected to be the dominant ones at this order. The third one, an EW correction to the LO interference term of $\order{\alphas\alpha^5}$,
despite its negligible size, must be included to guarantee IR safety, since it cancels divergences stemming from virtual corrections.

\begin{figure}
\centering
\begin{subfigure}{0.3\textwidth}
\captionsetup{skip=0pt}
\caption{}
\centering
\includegraphics[page=1,width=1.\linewidth]{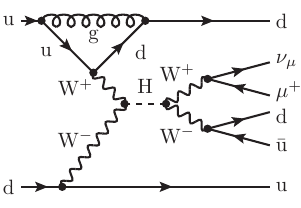}
\end{subfigure}\quad
\begin{subfigure}{0.3\textwidth}
\captionsetup{skip=0pt}
\caption{}
\centering
\includegraphics[page=2,width=1.\linewidth]{diagrams/diagrams.pdf}
\end{subfigure}\quad
\begin{subfigure}{0.3\textwidth}
\captionsetup{skip=0pt}
\caption{}
\centering
\includegraphics[page=3,width=1.\linewidth]{diagrams/diagrams.pdf}
\end{subfigure}
\caption{Examples of virtual diagrams of $\order{g_\mathrm s^2 g^6}$
  for a partonic process of type $\Pq\Pq \to \Pl^+\nu_\Pl 4\Pq$,
  explicitly $\Pu\Pd \to \mu^+ \nu_\mu \Pu\bar\Pu \Pd\Pd$. Diagram (a)
  contains only QCD singularities, diagram (b) only QED
  singularities, and diagram (c) mixed QED and QCD
  singularities.}
\label{fig:virt_qcd}
\end{figure}

The virtual contributions of $\order{\alphas\alpha^6}$ consist of
three different types. They are either interferences of diagrams of
$\order{g_\mathrm s g^7} \times \order{g_\mathrm s g^5}$, of
$\order{g^8} \times \order{g_\mathrm s^2 g^4}$, or of
$\order{g_\mathrm s^2 g^6} \times \order{g^6}$. The odd power of
$g_\mathrm s$ in both interfering diagrams of the first type indicates the
presence of an external gluon, whereas the gluon appears as an internal 
particle in the second and third types. Hence, in the first type,
only EW loops and QED singularities are present. Similarly, in the
second type, virtual EW loops are interfered with a QCD tree-level
diagram, which also leads to pure QED singularities. However, there
are diagrams of $\order{g_\mathrm s^2 g^6}$ in the third type that
include only QED singularities, others that include only QCD
singularities, and diagrams with singularities of both types (see
\reffi{fig:virt_qcd} for illustration). 
As a consequence the singularities of QED and QCD origin cannot be
separated based on subsets of diagrams.  They are only cancelled upon
including both QCD and QED real-emission contributions.

The subset of processes considered in our calculation is, like in the case of the EW
corrections, chosen to guarantee IR safety. A discussion of the neglected contributions
is given in \refse{se:neglected}.

\subsection{Neglected contributions}
\label{se:neglected}
\subsubsection{Misidentified jets}
\label{se:misidentification}
The contribution of misidentified jets is an irreducible background
for the considered process in all partonic channels.
It may happen that a
single jet or a jet pair has an invariant mass accidentally close to the $\PW$ or $\PZ$
mass and is thus wrongly attributed to the decay product of an EW boson.
There is, however, a
separable class of partonic processes that contributes only to the
physical process \refeq{eq:phys_proc_boo} at LO, namely the class
$\qaloproc$ in \refeq{eq:lo_subproc_aq}.
While the amount of partons in the final
state is too small to contribute to the process
\refeq{eq:phys_proc_res}, the photon can recombine with one of the three
partons into a massive jet. The very same behaviour can occur for
 all classes of processes with three partons and
two photons in the final state, \ie
$\Pq\gamma \to \Pl^+\nu_\ell 3\Pq 2\gamma$ at both $\order{\alpha^7}$
and $\order{\alphas\alpha^6}$ as well as $\Pq\Pg \to\Pl^+\nu_\ell 3\Pq
2\gamma$ and $\Pq\Pq \to \Pl^+\nu_\ell 2\Pq \Pg 2 \gamma$ at
$\order{\alphas\alpha^6}$.

The impact of these channels is tiny. We calculated the LO
contribution of $\Pq \gamma \to \Pl^+\nu_\ell 3\Pq \gamma$ and found
it to be below the per-mille level compared to the dominant
contribution of $\Pq\Pq \to \Pl^+\nu_\ell 4\Pq$. Although the
existence of a second photon at NLO increases the possibility for a
misidentification, we do not expect this effect to offset the
suppression by the additional power of the coupling constant to such
an extent that a significant NLO correction results.

\subsubsection{Photon-induced processes}
\label{se:ph-induced}
There are three classes of photon-induced processes: photon--photon-induced, photon--gluon-induced, and photon--quark-induced ones. 

The first class already appears at LO and receives both EW and QCD
corrections. Because of the strong dominance of quark PDFs over photon
PDFs, it is doubly suppressed compared to the dominant
contributions. We calculated the partonic processes belonging to this
class at LO. Their cross section relative to the quark-induced LO process is in both setups
presented in \refse{se:cuts}
at the level of a few parts per million and therefore irrelevant.
We did not calculate its NLO corrections
since no significance is expected.

The photon--gluon-induced channels appear as NLO contributions of
$\order{\alphas\alpha^6}$. In this class, the suppression due to the
photon PDF could possibly be balanced by an enhancement caused by the gluon PDF.
We performed an explicit calculation at LO and NLO and found their contribution to be
at the level of $0.1\%$ compared to the LO result shown in \refse{se:results} in both setups.

\begin{figure}
\centering
\includegraphics[page=4,width=.3\linewidth]{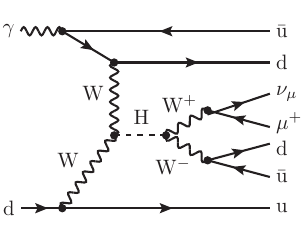}
\caption{Example of a photon-induced diagram with a VBS topology.}
\label{fig:ph-ind_QED}
\end{figure}

The photon--quark induced processes, which are already present at LO,
have a different nature
at $\order{\alpha^7}$ and at $\order{\alphas\alpha^6}$. While nothing
special happens for their QCD corrections,
at the $\order{\alpha^7}$
many new partonic processes with VBS topologies open up in
the photon-quark-induced channels.
An example diagram is shown in \reffi{fig:ph-ind_QED}. 
In the past, sizeable EW corrections stemming from this type of
process have been found for the scattering of $\PW^+\PW^-$ bosons \cite{Denner:2022pwc}. 
We performed a full calculation for the EW corrections
in both setups and found this type of processes to 
contribute $1.1\%$ and $1.3\%$ to the fiducial cross section in the
resolved and boosted setup, respectively. For the
$\order{\alphas\alpha^6}$ corrections,
we calculated the dominant partonic processes $\Pu \gamma/\Pc\gamma \to \ell\nu 5\Pq$
and found a contribution at the parts per million level.

At last, we calculated the processes $\Pu \gamma/\Pc\gamma \to \ell\nu 3\Pq\Pg (+\gamma)$ 
at LO and NLO. They contribute $0.2\%$ of the cross section in both setups. 
Taking all partonic processes into account, we expect a contribution of
less than $1\%$. 

\subsection{Details and validation of the calculation}
\label{se:validation}
The results presented here were produced using the parton-level Monte
Carlo program \mocanlo~\cite{Denner:2026phn,denner_2026_19829093}, in which the subtraction
of IR divergences in the real radiation is realised with the
Catani--Seymour dipole formalism
\cite{Catani:1996vz,Dittmaier:1999mb,Catani:2002hc,Dittmaier:2008md}.
All LO and NLO matrix elements are obtained from \recola
\cite{Actis:2012qn} and \collier \cite{Denner:2016kdg}.  \mocanlo
employs multichannel importance sampling for the Monte Carlo
integration \cite{Berends:1994pv,Denner:1999gp,Dittmaier:2002ap}, a
necessary feature for the computation of high-multiplicity processes
with complex resonance structures. More precisely, the contributions
of each partonic process are separately calculated, and dedicated
integration channels are constructed based on their resonance
structure. In the particular case of semileptonic VBS, the larger
amount of partonic processes compared with 
fully-leptonic VBS represents an additional challenge. To
tackle it, a new feature was implemented in \mocanlo, called
\emph{final-state merging}.  This allows the simultaneous integration of
multiple partonic processes with the same initial state and
with final states that differ only in the flavours of the 
fermions.  To this end, the integration channels of all
simultaneously integrated partonic processes are generated, and
duplicates are discarded.  Table \ref{tab:fs_merging_improvement}
summarises the reduction of independent calculations that was achieved
with \mocanlo's final-state merging. For the considered processes, it
reduces the amount of required calculations by one order of magnitude.

\begin{table}[t]
  \centering
  \begin{tabular}{l|cc}
    \hline
    \rowcolor{blue!30}
    \multicolumn{1}{c}{class}
    &\multicolumn{1}{c}{partonic processes}
    &\multicolumn{1}{c}{required calculations}\\
    \hline
    $\Pq\Pq   \to \Pl^+\nu_\ell \,  4 \Pq \gamma$  & 202 & 32 \\ 
    $\Pq\Pq   \to \Pl^+\nu_\ell \,  4 \Pq \Pg$         & 202 & 32 \\
    $\Pq\Pg   \to \Pl^+\nu_\ell \,  5 \Pq$                & 96 & 8 \\
    \hline
  \end{tabular}
  \caption{Classes of calculated partonic processes. The second
    and third columns display the size of the class and the  
    number of separate calculations that were necessary when using
    \mocanlo's final-state merging feature, respectively. 
Note that the number of partonic processes that need to be calculated
without final-state merging is about twice as big as the numbers
shown in the left column, since processes with reflected initial states need to be
calculated separately. On the other hand, merging of partonic processes that are
related by swapping  quarks of the first and second generation
(PDF merging) already reduces the number of required calculations by a
factor two. 
    }
        \label{tab:fs_merging_improvement}
\end{table}
All results presented here were independently calculated using the
in-house Monte Carlo program \bbmc. Although both codes rely on the
use of \recola and \collier for the calculation of the matrix
elements, the generation and evaluation of integration channels, the
final-state merging, the recombination and event selection routine, as
well as the Catani--Seymour subtraction are implemented
independently. We verified that the fiducial cross sections agree
between the two codes within a level of $3\sigma$ of the integration
error. The differential
distributions were checked on a bin-by-bin basis. The amount of bins
with deviations larger than $3\sigma$ follows the expected Gaussian
statistics. Furthermore, we checked for a selection of
partonic channels the agreement of the Catani--Seymour subtraction
 on the phase-space-point level, where we did not
find any significant discrepancy. Moreover, we calculated the
dominant contribution at $\order{\alphas\alpha^6}$ by varying the
$\alpha_{\rm dipole}$~parameter, which is a technical parameter entering the
evaluation of integrated and subtraction dipoles in the Catani--Seymour algorithm (see \citere{Nagy:2003tz}),
between $\alpha_{\rm dipole} =10^0$ and $\alpha_{\rm dipole} =
10^{-2}$ and found agreement of both codes. For
the check of the recombination and event selection, we generated up to
$\order{10^6}$ phase-space points with \bbmc and fed them to \mocanlo for evaluation
and vice versa. We observed, up to different technical
cuts, perfect agreement for the accepted and rejected phase-space points. 

\section{Setup of the calculation}
\label{se:setup}
\subsection{Input parameters}
\label{se:inputs}  

The calculations are done for the LHC with a centre-of-mass (CM)
energy of $13.6\TeV$.

We use the values from the PDG review 2024~\cite{ParticleDataGroup:2024cfk}
for the masses and widths of the massive gauge bosons,
\begin{subequations}
\begin{equation}
\begin{aligned}
                \MZ &=  91.1880\GeV,       & \quad \quad \quad \GZ &= 2.4955\GeV, \\
                \MW &=  80.3692\GeV,       & \GW &= 2.085\GeV,
\end{aligned}
\end{equation}
as well as for the masses of the top quark and the Higgs boson,
\begin{equation}
\begin{aligned}
                  \Mt   &=  172.6\GeV,      & \quad \quad \quad \Gt &= 0 \GeV, \\
                M_{\rm H} &=  125.2\GeV,    &  \GH  &=  4.1 \times 10^{-3}\GeV.
\end{aligned}
\end{equation}
\label{eqn:ParticleMassesAndWidths}
\end{subequations}
On the other hand, the Higgs-boson width reported above is taken from \citere{Heinemeyer:2013tqa},
while the top-quark width is set to zero. 
Since we assume a diagonal CKM matrix and neglect all contributions
with external bottom quarks, there are no resonant top quarks in the
considered processes. Finally, all leptons and light quarks, including the bottom
quark, are treated as massless.

We use the $G_\mu$ scheme \cite{Denner:2000bj} to define the
electromagnetic coupling, which fixes the EW coupling $\alpha$ using the Fermi
constant 
\begin{equation}
\GF    = 1.16638\times 10^{-5}\GeV^{-2}
\end{equation}
and the complex masses
\begin{equation}
\mu_\PZ^2 = \MZ^2 - \ri\MZ\GZ,
\qquad
\mu_\PW^2 = \MW^2 - \ri\MW\GW
\end{equation}
as input parameters via 
\begin{equation}
  \alpha = \frac{\sqrt{2}}{\pi} G_\mu \left|\,\mu_\PW^2\left( 1 -
      \frac{\mu_\PW^2}{\mu_\PZ^2} \right)\right| 
 = 1/132.0360732365\,.
\end{equation}

{\sloppy
We employ the NNLO QED PDF set NNPDF4.0 (\texttt{NNPDF40\_nnlo\_as\_01180\_qed})
with $\alphas(\MZ) = 0.118$ \cite{NNPDF:2024djq}
via LHAPDF \cite{Andersen:2014efa,Buckley:2014ana} in the
$N_\text{F}=5$ fixed-flavour scheme in both the LO and NLO predictions.
Initial-state collinear singularities are absorbed by ${\overline{\rm
    MS}}$ redefinition of the PDFs.
}

The central renormalisation and factorisation scales are chosen as the
geometric average of the transverse momenta of the tag jets
\begin{equation}
\label{eq:defscale}
\mu_{\rm ren}^{\rm central} = \mu_{\rm fac}^{\rm central} = \sqrt{p_{\rm T, j_1}\, p_{\rm T, j_2}},
\end{equation}
where the precise definition of the tag jets $\Pj_1$ and $\Pj_2$ 
is given below in \refse{se:cuts}. Based on
this central scale, we perform a 7-point scale variation of both the
renormalisation and factorisation scale, \ie we calculate the
observables for the pairs
\begin{equation}
(\mu_\mathrm{ren}/\mu_{\rm ren}^{\rm central},\mu_\mathrm{fact}/\mu_{\rm fact}^{\rm central}) = (0.5,0.5),
(0.5,1),(1,0.5),(1,1),(1,2),(2,1),(2,2)
\end{equation}
of renormalisation and factorisation scales
and use the resulting envelope to estimate the perturbative (QCD) uncertainty.
Since the $\mathcal{O}(\alpha^6)$ and $\mathcal{O}(\alpha^7)$ only depend on the factorisation scale,
the scale envelope is effectively computed over a 3-point scale variation.

All partonic channels with external
bottom quarks are ignored. For final-state bottom quarks, omitting these channels is
equivalent to assuming a perfect bottom-jet veto. 
Moreover, partonic channels with initial-state $\Pb\bar\Pb$ pairs are negligibly
small at LO \cite{Denner:2024xul}, and we do not expect them to be relevant
at NLO. 

\subsection{Event selection}
\label{se:cuts}
The event selection used for this analysis is inspired by the CMS and
ATLAS measurements of VBS into semileptonic final states
\cite{ATLAS:2018tav,CMS:2021qzzv1}. It closely follows the selection
criteria used for the LO calculation of this process \cite{Denner:2024xul}
with an additional treatment of real photons and a separation cut
of jets and antimuons.

The rapidity $y$ and the transverse momentum $p_{\rm T}$ of a
particle are defined as
\begin{equation}
y = \frac 12 \ln \frac{E + p_z}{E-p_z}, \qquad 
p_{\rm T} = \sqrt{p_x^2 + p_y^2}, 
\end{equation}
where $E$ is the energy of the particle, $p_z$ the component of its momentum along
the beam axis, and $p_x,p_y$ the components perpendicular to the beam axis.
Only partons with rapidity $|y| < 5$ are considered for recombination,
while particles with larger $|y|$ are assumed to be lost in the beam
pipe. We perform a two step recombination procedure. In the first 
step, the Cambridge--Aachen algorithm \cite{Dokshitzer:1997in,Wobisch:1998wt}
is used with a resolution parameter $R = 0.1$ to dress the
antimuons with photons that emerge as NLO real radiation. 

The dressed antimuon has to fulfil
\begin{align}
p_{\rm T,\mu} > 30\GeV, \qquad |y_\mu| < 2.4,
\end{align}
while the missing transverse momentum is required to satisfy
\begin{align}
p_{\rm T, \rm{mis}} > 30 \GeV
\end{align}
and is computed as the transverse part of the neutrino momentum $p_{\nu_\mu}$ at Monte Carlo-truth
level. Furthermore, the transverse mass of the $\PW$ boson is bounded by
\begin{align}
M_{\rm T,\PW} = \sqrt{2p_{\rm T,\mu} p_{\rm T,\nu_\mu}(1 - \cos \Delta \phi_{\mu\nu_\mu})} < 185\GeV
\end{align}
with the azimuthal-angle difference $\Delta \phi_{ij} = \min(|\phi_i -
\phi_j|, 2\pi - |\phi_i - \phi_j|)$.

In a second recombination step, the QCD partons are clustered with
each other and the remaining photons with the anti-$k_\text{T}$
algorithm \cite{Cacciari:2008gp}. The clustering is performed independently
twice: once with resolution parameter $R =0.4$ and once with $R = 0.8$.
The resulting jets are called AK4 and AK8 jets, respectively. Note that AK8 jets
are a particular realisation of the fat jets $\PJ$ introduced in the previous section [see
\refeq{eq:phys_proc_boo}]. 

On the combined set of AK4 and AK8 jets, we apply different particle selection
criteria. AK4 jets that do not fulfil the conditions
\begin{equation}
p_{\rm T,\Pj_{\rm{AK4}}} > 30\GeV, \qquad |y_{\Pj_{\rm{AK4}}}| < 4.7
\end{equation}
are discarded. Similarly, AK8 jets are excluded if they do not respect
\begin{align}
p_{\rm T, \Pj_{\rm{AK8}}} > 200\GeV, \qquad |y_{\Pj_{\rm{AK8}}}| < 2.4, \qquad 40\GeV < M_{\Pj_{\rm{AK8}}} < 250 \GeV\,.
\end{align}
To remove most of the overlap between the remaining AK4 and AK8 jets, we demand
\begin{equation}
\Delta R_{\Pj_{\rm{AK4}}\Pj_{\rm{AK8}}} > 0.8
\end{equation}
for every AK4 jet and discard it otherwise. As usual,
the distance $\Delta R_{ij}$ for two objects $i$ and $j$ is defined as
\begin{equation}
\Delta R_{ij} = \sqrt{(\Delta \phi_{ij})^2 + (\Delta y_{ij})^2}
\end{equation}
with the rapidity difference $\Delta y_{ij} = |y_i - y_j|$.

We consider two setups based on the multiplicity of the resulting AK8 jets. 
We call the first case \textit{resolved setup}, where we require the absence of any AK8 jet
and simultaneously the presence of at least four AK4 jets. We call the second case
\textit{boosted setup}, where exactly one AK8 jet and at least two AK4 jets are found. 

In both setups, the two AK4 jets with highest invariant mass are called \textit{tag jets}
and must obey 
\begin{equation}\label{eq:cut_tag_jets}
M_{\Pj_1\Pj_2} > 500\GeV, \qquad \Delta y_{\Pj_1 \Pj_2} > 2.5.
\end{equation}
Furthermore, the tag jet with largest transverse momentum, called
leading tag jet in the following, must satisfy
\begin{equation}
p_{\rm T,\Pj_1} > 50\GeV.
\end{equation}
All AK4 jets (and the AK8 jet in the boosted setup) must have a minimal separation
from the antimuon, namely
\begin{equation}\label{eq:cutRjmu}
\Delta R_{\Pj_{\rm{AK4}}\mu} > 0.1 \quad \mathrm{and} \quad \Delta R_{\Pj_{\rm{AK8}}\mu} > 0.1.
\end{equation}
In the resolved setup, we define the mass $M_V$ of the hadronically-decaying
vector boson as the invariant mass of the two AK4
jets that are not tag jets and whose invariant mass is closest to
$85\GeV$. These jets are denoted as \emph{decay jets} in the following.
However, in the boosted setup, $M_V$ is the invariant mass of the single AK8 jet.
We require
\begin{equation}
65\GeV < M_V < 105\GeV\ \quad \mathrm{and} \quad 70\GeV < M_V < 115\GeV
\end{equation}
in the resolved and boosted setup, respectively.

\subsection*{On IR safety and a minimal distance between jets and antimuons}
The experimental setups \cite{ATLAS:2018tav,CMS:2021qzzv1} that inspired
our theoretical study both at LO and at NLO do not specify a distance
cut between jets and antimuons. While such a cut is not necessary at LO,
the matrix element of the process involves an unregulated collinear singularity at NLO
if such a cut is not applied. In the resolved setup, this singularity
is avoided by requiring the separation cut \refeq{eq:cutRjmu}.

\begin{figure}
\centering
\includegraphics[page=5,width=.3\linewidth]{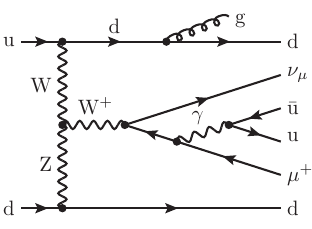}
\caption{Real radiation diagram with a potential triple collinear singularity.}
\label{fig:real-triple-collinear}
\end{figure}

However, in the boosted setup, the situation is more complicated, since we
demand only three out of the four jets potentially present at LO.
Thus,
a triple collinear configuration can occur within the
fiducial volume, \ie if a quark, an antiquark of the same flavour, and the antimuon become
collinear. This, for instance, appears if the antimuon emits a photon that
splits into a quark--antiquark pair, as illustrated in \reffi{fig:real-triple-collinear}.  
This
singularity is not taken care of by the NLO Catani--Seymour formalism
and appears both in the real contribution and in the integrated
subtraction term cancelling the singularity of the final-state $\gamma\to\Pq\bar\Pq$ splitting.  This
singularity is actually present in the boosted setup defined in
\refse{se:cuts}. There, it may happen that the quark and antiquark
that are collinear to the antimuon are recombined with a third parton
to form an AK8 jet that passes the selection cuts.

In practice, in the Monte Carlo integration the singularity is
captured by technical cuts and its effect is negligible in the boosted
setup of \refse{se:cuts}.  To estimate
the effect, we introduced a technical cut on the ratio of the
invariant masses of any pair of external particles $s_{ij}$ and the
partonic CM energy $\hat s$, leading to terms proportional
to $\log (s_{ij,\mathrm{min}}/\hat s)$. We varied this cut in the interval
$10^{-10} < s_{ij,\mathrm{min}}/\hat s < 10^{-6}$ for the dominant
partonic channel in the boosted setup and found an effect at the
sub-per-mille level relative to the Born cross section.

A proper treatment of this singularity would require to introduce a separation cut
$\Delta R_{\Pj_{\rm{AK8}}\mu}> 0.8$ between the AK8 jet and the antimuon. Then, all triple collinear phase-space points would be removed. 
Actually, in more recent publications by ATLAS \cite{ATLAS:2025omi}
and CMS \cite{CMS:2021qzzv2} such a cut is employed.

\section{Results}
\label{se:results}
\subsection{Fiducial cross sections}
\label{se:crosssection}

In \refta{tab:integrated_xsec} we present the fiducial cross
sections at LO, \ie $\order{\alpha^6}$, and the corresponding NLO
corrections of $\order{\alpha^7}$ and $\order{\alphas\alpha^6}$ in the
resolved and boosted setups defined in \refse{se:cuts}. 
\begin{table}
\centering
\begin{tabular}{c|cc}
\hline
\rowcolor{blue!30}
setup & resolved & boosted \\
\hline
\rule[-1.4ex]{0pt}{4.0ex} 
$\sigma_{\rm{LO}}^{\alpha^6}$ [fb]
 & $10.2267(3)$ & $2.8872(1)$ \\ \hline
\rule[-1.4ex]{0pt}{4.0ex} 
$\Delta\sigma_{\rm{NLO}}^{\alpha^7}$ [fb]
 & $-1.428(5)$ & $-0.724(3)$ \\
\rule[-1.4ex]{0pt}{4.0ex} 
$\delta^{\mathrm{EW}}=\Delta\sigma_{\rm{NLO}}^{\alpha^7}/\sigma_{\rm{LO}}^{\alpha^6}$ [\%]     &
 $-13.9$     & $-25.0$ \\ \hline
\rule[-1.4ex]{0pt}{4.0ex} 
$\Delta\sigma_{\rm{NLO}}^{\alphas\alpha^6}$ [fb]
 & $-1.58(1)$  & $0.133(5)$ \\
\rule[-1.4ex]{0pt}{4.0ex} 
$\delta^{\mathrm{QCD}}=\Delta\sigma_{\rm{NLO}}^{\alphas\alpha^6}/\sigma_{\rm{LO}}^{\alpha^6}$ [\%]
 & $-15.4$     & $4.6$ \\ \hline
 
\end{tabular}
\caption{Integrated cross sections for $\order{\alpha^6}$, $\order{\alpha^7}$, and $\order{\alphas\alpha^6}$ in fb and relative to the LO EW contribution.}
\label{tab:integrated_xsec}
\end{table}
Compared to the setups considered in
\citere{Denner:2024xul}, we introduced the additional distance cut
\refeq {eq:cutRjmu} between the jets and the antimuon and used the
higher CM energy of LHC Run 3 of $13.6\TeV$. We observe that the fiducial cross
sections rise from $9.02\fb$ and $2.51\fb$ at $\sqrt s = 13\TeV$ to
$10.23\fb$ and $2.89\fb$ at $\sqrt s = 13.6\TeV$ for the resolved and
boosted setup, respectively. We checked that the
additional distance cut does not affect the LO result significantly for the
dominant partonic processes $\Pu\Pu \to \ell^+\nu_\ell 4\Pq$, which
contribute roughly one third of the total cross section.

The EW corrections are, as expected, large and negative, and amount to 
$-14\%$ in the resolved and $-25\%$ in the boosted setup. These sizeable
corrections can be traced back to the presence of Sudakov logarithms
stemming from virtual contributions with massive particles in the
loops. Based on \citeres{Denner:2000jv,Accomando:2006hq}, the
angular-independent leading-logarithmic (LL) corrections for the
scattering of transverse vector bosons read 
\begin{align}\label{eq:LL}
  \delta_\mathrm{LL}^\mathrm{EW} = \frac{\alpha}{4\pi}
  \left[-4C_W^\mathrm{EW}\log^2\left(\frac{Q^2}{M_\PW^2}\right) + 2
    b_W^\mathrm{EW}\log\left(\frac{Q^2}{M_\PW^2}\right)\right],
\end{align}
where $Q$ is the characteristic energy scale of the VBS subprocess, which is given by
the CM energy of the two vector bosons. The constants
$C_W^\mathrm{EW} = 2/s_\mathrm w^2$ and $b_W^\mathrm{EW} =
19/(6s_\mathrm w^2)$ depend on the sine of the weak mixing angle
$s_\mathrm w$. The result \refeq{eq:LL} is independent of the specific VBS
process and is expected to capture the dominant corrections
\cite{Biedermann:2016yds,Denner:2019tmn,Denner:2020zit} for all of them.
Thus, as long as the VBS diagrams dominate, it should also apply to our case, where the
hadronically decaying vector boson can be either a W or a Z boson in
the considered processes \refeq{eq:phys_proc_res} and
\refeq{eq:phys_proc_boo}.

In the case of semileptonic VBS, the scale $Q$ corresponds to the
invariant mass of the two leptons and either the two non-tagged AK4
jets or the AK8 jet, depending on the setup, since these objects are
identified as the decay products of the two vector bosons.
The dominance of the VBS processes, which is a precondition for \refeq{eq:LL}
to be valid, can be seen as follows.
First, the contribution of the resonant Higgs boson to the fiducial
cross section was determined in \citere{Denner:2024xul} based on a
pole approximation for the Higgs boson and found to be at the level of
$3\%$, which is negligible for a LL estimate.
Second, the contribution of non-VBS diagrams can be estimated from
partonic processes that do not receive a  VBS contribution.
These are characterised by different generations for the initial- and final-state
quarks. All these processes contribute to triple
vector-boson production, where one of the vector bosons is not allowed
to become resonant because of the invariant mass cut \refeq{eq:cut_tag_jets}
on the tag jets.  The sum of these contributions has been found to
be at the level of $0.1\permil$ 
in both setups at LO.\footnote{Note
that these types of contributions are expected to have the largest
relative QCD corrections. This behaviour is already known from fully
leptonic VBS processes \cite{Denner:2020zit}. The presence of an additional jet in the
final state from initial-state radiation helps to evade the strict
invariant-mass cut on the two tag jets, which do not have to be
produced via an $s$-channel resonance any more. Then, the
additional resonance enhances the correction.} 

The average invariant masses of the four/three reconstructed objects,
inferred from the differential distribution shown in \reffi{fig:distr_M_VV}, turn out to be 
\begin{align}\label{eq:average_energy}
\langle M_{\ell^+\nu_\ell\Pj\Pj} \rangle = 372.5\GeV, \qquad \langle M_{\ell^+\nu_\ell\PJ} \rangle = 675.5 \GeV
\end{align}
in the two setups. Using these average values, the Sudakov
approximation \refeq{eq:LL} predicts a relative EW correction of
\begin{align}
\delta_\mathrm{LL}^\mathrm{EW} = -15.1\%, \qquad \delta_\mathrm{LL}^\mathrm{EW} = -31.9\%
\end{align}
in the resolved and boosted setup, respectively.
A more refined estimate is obtained by applying the Sudakov
approximation on a bin-by-bin basis to the differential distribution
in the invariant mass of the two-vector-boson system and then averaging the bin-by-bin estimates,
leading to
\begin{align}
\delta_\mathrm{LL}^\mathrm{EW} = -14.3\%, \qquad \delta_\mathrm{LL}^\mathrm{EW} = -29.5\%, \label{eq:Sudakov_differential}
\end{align}
which is within a few per cent around the full relative $\order{\alpha}$ correction
in \refta{tab:integrated_xsec}. 

The QCD corrections turn out to be negative in the resolved and positive in the boosted setup,
amounting to $-15.4\%$ and $+4.6\%$, respectively. Both values are in line with the expectations
for QCD corrections. The total NLO corrections thus reach approximately $-30\%$ and $-20\%$ in the two setups.

\subsection*{Scale variations}
\begin{table}
\centering
\renewcommand{\tabcolsep}{5pt}
\begin{tabular}{c|cccccc}
\hline
\rowcolor{blue!30}
setup & \multicolumn{3}{c}{resolved} & \multicolumn{3}{c}{boosted} \\
      & min & central & max          & min & central & max \\\hline 
$\sigma_{\rm{LO}}^{\alpha^6}$ [fb]                                              & 
    $9.4479(2)$ & $10.2267(3)$ & $11.1342(3)$   & $2.6128(1)$  & $2.8872(1)$& $3.2173(1)$ \\ 
$\sigma_{\rm{LO}}^{\alpha^6} + \Delta \sigma_{\rm{NLO}}^{\alpha^7}$ [fb]        & 
    $8.157(5)$  & $8.798(5)$   & $9.542(6)$     & $1.967(2)$  & $2.162(3)$  & $2.396(3)$  \\
$\sigma_{\rm{LO}}^{\alpha^6} + \Delta \sigma_{\rm{NLO}}^{\alphas\alpha^6}$ [fb] & 
    $8.37(1)$   & $8.64(1)$    & $8.77(1)$      & $2.981(4)$ & $3.020(5)$   & $3.074(5)$  \\
$\sigma_{\rm{NLO}}^{\mathrm{tot}}$ [fb]                                         & 
    $6.78(1)$   & $7.21(1)$    & $7.46(1)$      & $2.247(7)$  & $2.296(6)$ & $2.335(5)$  \\
\end{tabular}
\caption{Scale variation of fiducial cross sections.}
\label{tab:scale_variations}
\end{table}

Since we are only considering the EW contribution at LO, there is no
dependence on the renormalisation scale, but only on the factorisation scale
via the PDFs. The same is true for the EW corrections. The
variation of the renormalisation scale is thus only relevant for the
QCD corrections. The extrema of the 7-point scale variations are
presented in \refta{tab:scale_variations}. The inclusion of NLO QCD
and EW effects reduces the scale dependence from
${}^{+8.9\%}_{-7.6\%}$ at LO to ${}^{+3.5\%}_{-6.0\%}$ at NLO in the
resolved and from ${}^{+11.4\%}_{-9.5\%}$ to ${}^{+1.7\%}_{-2.1\%}$ in
the boosted setup.

\subsection{Differential distributions}
In this section we present the differential distributions of selected
observables. The plots always show the absolute differential cross
sections at LO (black), the NLO EW (red) and QCD (magenta) cross
sections separately, and the total NLO cross section (green).
Moreover, a dedicated panel illustrates 
the relative NLO EW and QCD corrections.
Each figure presents results for the two setups side by side, showing identical
or related observables, with the resolved and boosted cases on the left and right, respectively.
 Since the LO
behaviour has already been discussed in \citere{Denner:2024xul}, the
focus of this section is on the NLO corrections.

\begin{figure}
\centering
\begin{subfigure}{0.49\textwidth}
\centering
\includegraphics[page=14,width=1.\linewidth]{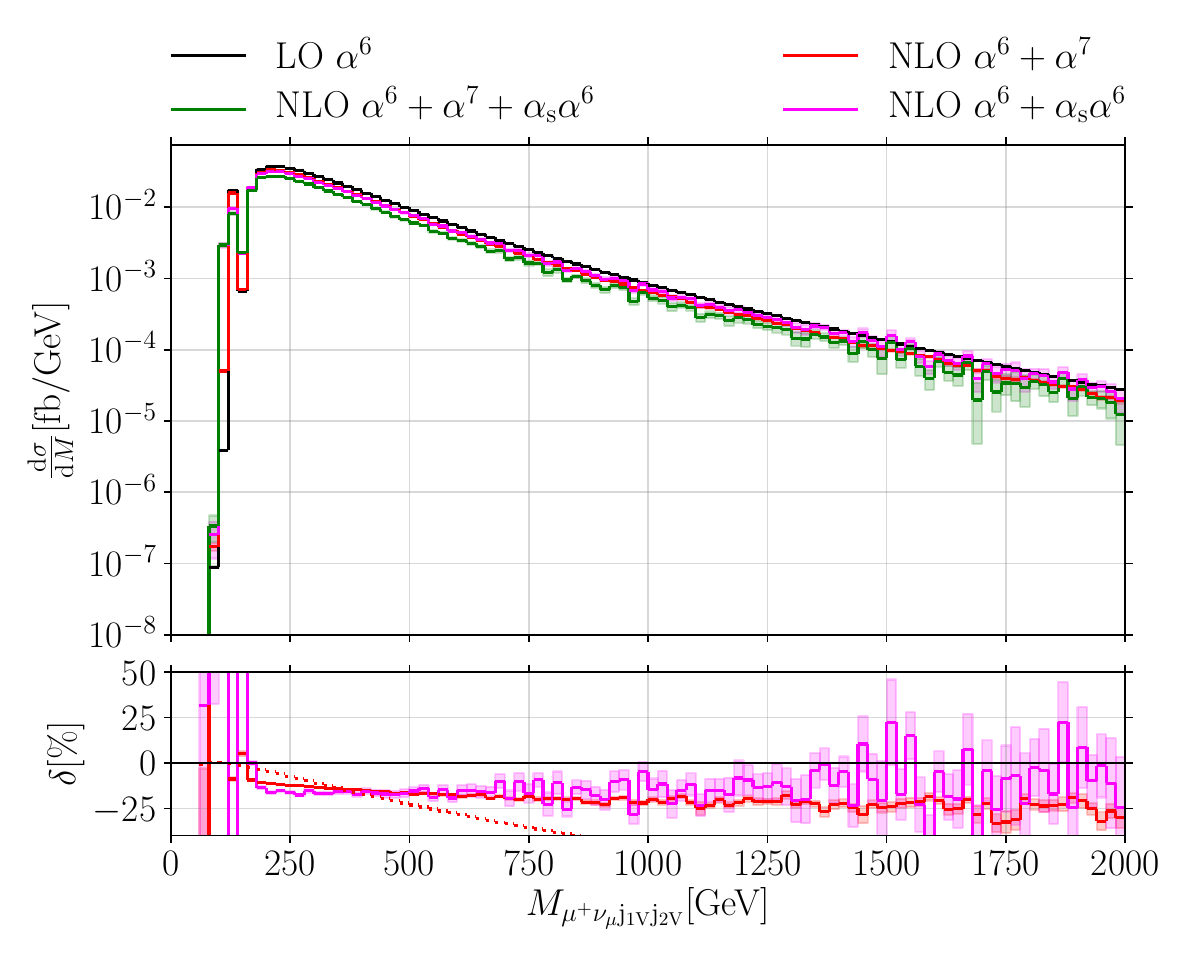}
\end{subfigure}
\hspace{\fill}
\begin{subfigure}{0.49\textwidth}
\centering
\includegraphics[page=14,width=1.\linewidth]{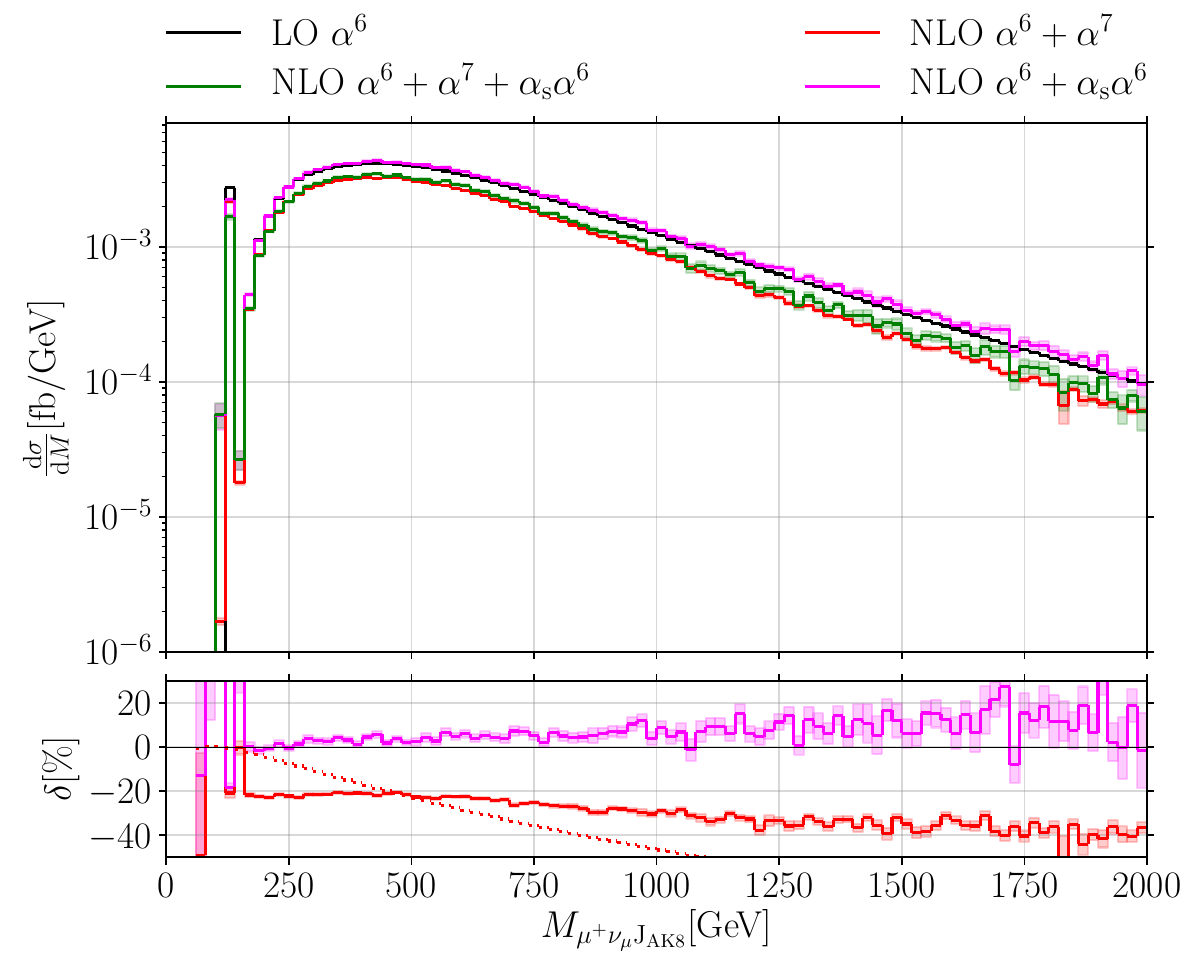}
\end{subfigure}
\caption{Differential cross section with respect to the invariant mass
  of the two-vector-boson system.   
The upper panel shows absolute cross sections, the lower panel the
relative corrections. Shaded bands denote the integration error.
} 
\label{fig:distr_M_VV}
\end{figure}
We first show the differential cross section with respect to the
invariant mass of the two scattering vector bosons
in \reffi{fig:distr_M_VV}. This observable can not be measured in experiments
owing to the presence of the neutrino as decay product of one of the
vector bosons. However, in a Monte Carlo integrator
the true momentum of the neutrino is accessible and the CM energy of the
two scattering vector bosons can be fully determined.
Above the Higgs-boson resonance, the
EW corrections are negative and follow the familiar decreasing pattern
towards larger energies. This behaviour can be traced back to the
appearance of Sudakov logarithms and is expected for energy-like observables.
We used this distribution in \refse{se:crosssection} to extract the differential
Sudakov approximation, whose results are given in
Eq.~\eqref{eq:Sudakov_differential}. Even if giving a fair estimate of NLO EW corrections at
the integrated level, the relative corrections
resulting from Eq.~\eqref{eq:LL} and shown in \reffi{fig:distr_M_VV}
as dashed red curve do not provide an adequate description of the exact NLO relative corrections
to the invariant-mass distribution. We attribute this to
angular-dependent leading logarithms of the form $\alpha\log({Q^2}/{M_\PW^2})
\log\left({t}/{s}\right)$, which are not included in
Eq.~\eqref{eq:LL} but have been shown to be sizeable
for invariant-mass distributions in other processes \cite{Pagani:2021vyk,Lindert:2023fcu,Denner:2024yut}.
In contrast, the QCD corrections are rather flat above $250\GeV$ in
the resolved setup and even increase in the boosted setup with the invariant 
mass. Around the Higgs-boson resonance, the Sudakov
approximation \refeq{eq:LL} is not valid, and we observe the effects
of radiative return like for the invariant-mass distribution of the
two decay jets (see \reffi{fig:distr_M_bos} below): the peak is
smeared out at NLO with huge positive QCD corrections in the
surrounding bins and negative ones in the bin containing the Higgs
resonance at LO. We note that also the EW corrections are, although
smaller, positive in the surrounding bins. 

In \reffi{fig:distr_M_bos} we present the distributions in the
invariant mass of the hadronically decaying vector boson, \ie of the
AK4 decay-jet pair in the resolved and of the unique AK8 jet in the
boosted setup.
\begin{figure}
\centering
\begin{subfigure}{0.49\textwidth}
\centering
\includegraphics[page=13,width=1.\linewidth]{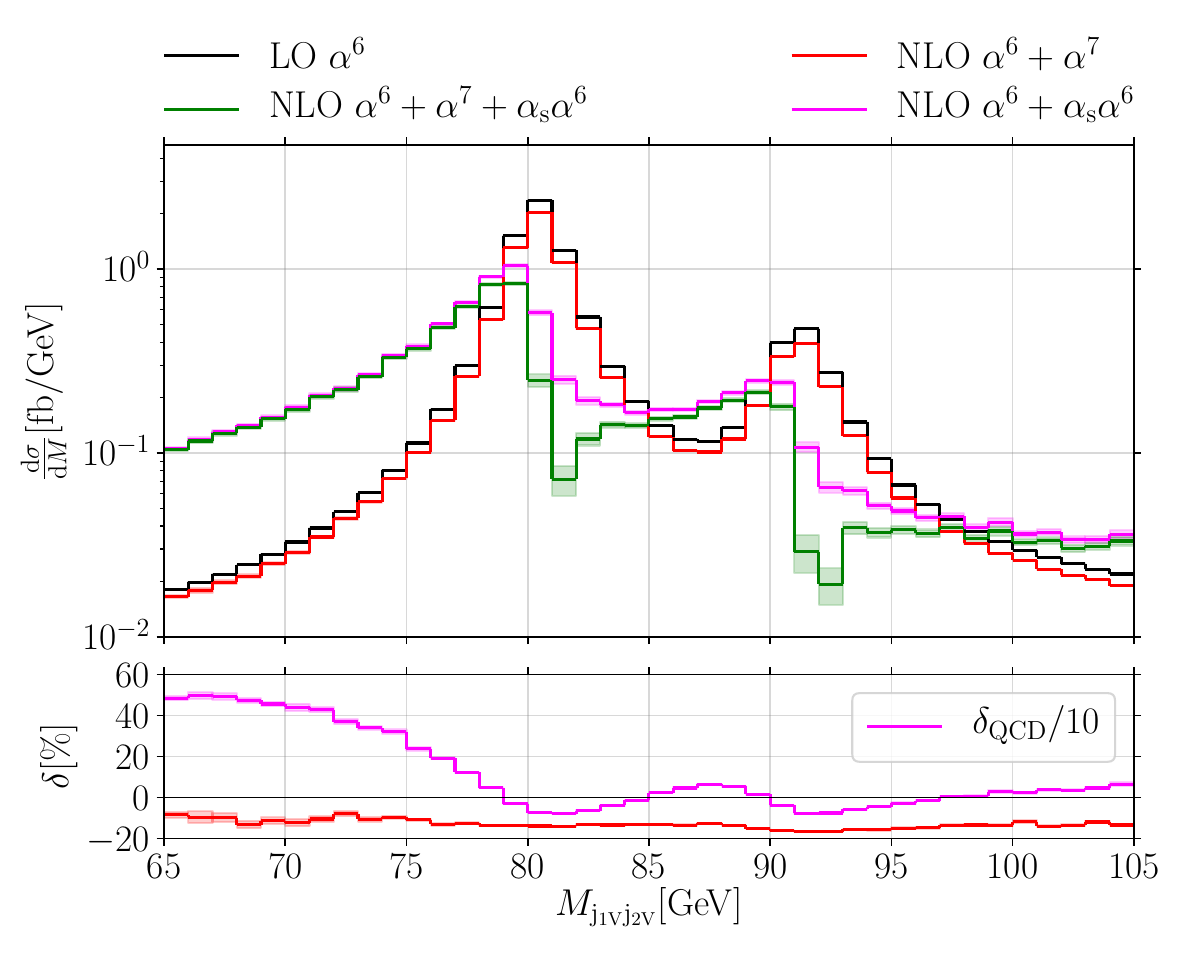}
\end{subfigure}
\hspace{\fill}
\begin{subfigure}{0.49\textwidth}
\centering
\includegraphics[page=13,width=1.\linewidth]{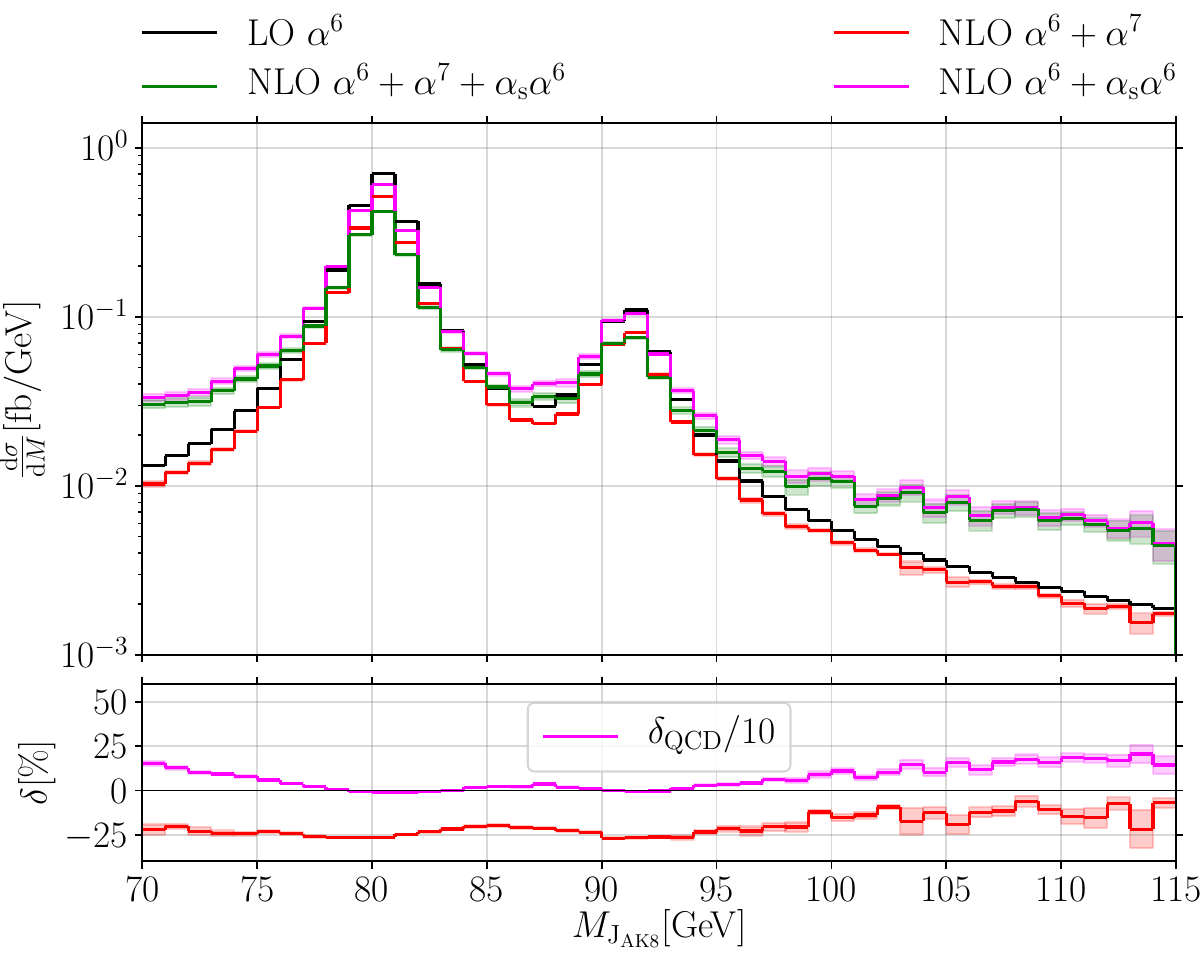}
\end{subfigure}
\caption{Differential cross section with respect to the invariant mass
  of the hadronically decaying vector boson. 
  Same structure as in \reffi{fig:distr_M_VV}.
The relative QCD corrections in the lower panel are scaled down by a factor of
$10$.
}
\label{fig:distr_M_bos}
\end{figure}
We observe a drastic change in the absolute cross section from
LO to NLO. The main effect stems from the QCD corrections. The two
resonance peaks of the W and Z~boson get smeared out and shifted towards
lower invariant masses. This effect is explained by the phenomenon of
radiative return. When two jets are produced via a
resonance and one of them emits a gluon that is not recombined with them, their
invariant mass becomes smaller than the resonance.  Thus, the QCD
corrections are negative at the LO peaks and turn positive for smaller
invariant masses.  A moderate increase of the QCD corrections is
observed between the LO peaks owing to radiative \PZ-boson decays with
unrecombined gluons. Below the $\PW$ mass, the QCD corrections become
huge, reaching up to $500\%$ in the resolved and $150\%$ in the
boosted setup. The opposite effect of radiative return takes place in
the region above the $\PZ$ resonance, especially in the boosted setup,
where the QCD corrections reach almost $200\%$. 
There, an AK8 jet can be produced via a resonance and get recombined with a
QCD emission from the initial state or one of the tag jets.  Note that we rescaled the relative QCD
corrections in \reffi{fig:distr_M_bos} by a factor 10.
The EW corrections follow a similar pattern, but less pronounced. They
remain negative in both setups, but are smaller in regions with
increased QCD corrections. Clearly, radiative return also exists for
the emission of a photon from the jets originating from the resonance, but it has to
compete with the negative Sudakov suppression, which is not present in
the QCD case.

\begin{figure}
\centering
\begin{subfigure}{0.49\textwidth}
\centering
\includegraphics[page=3,width=1.\linewidth]{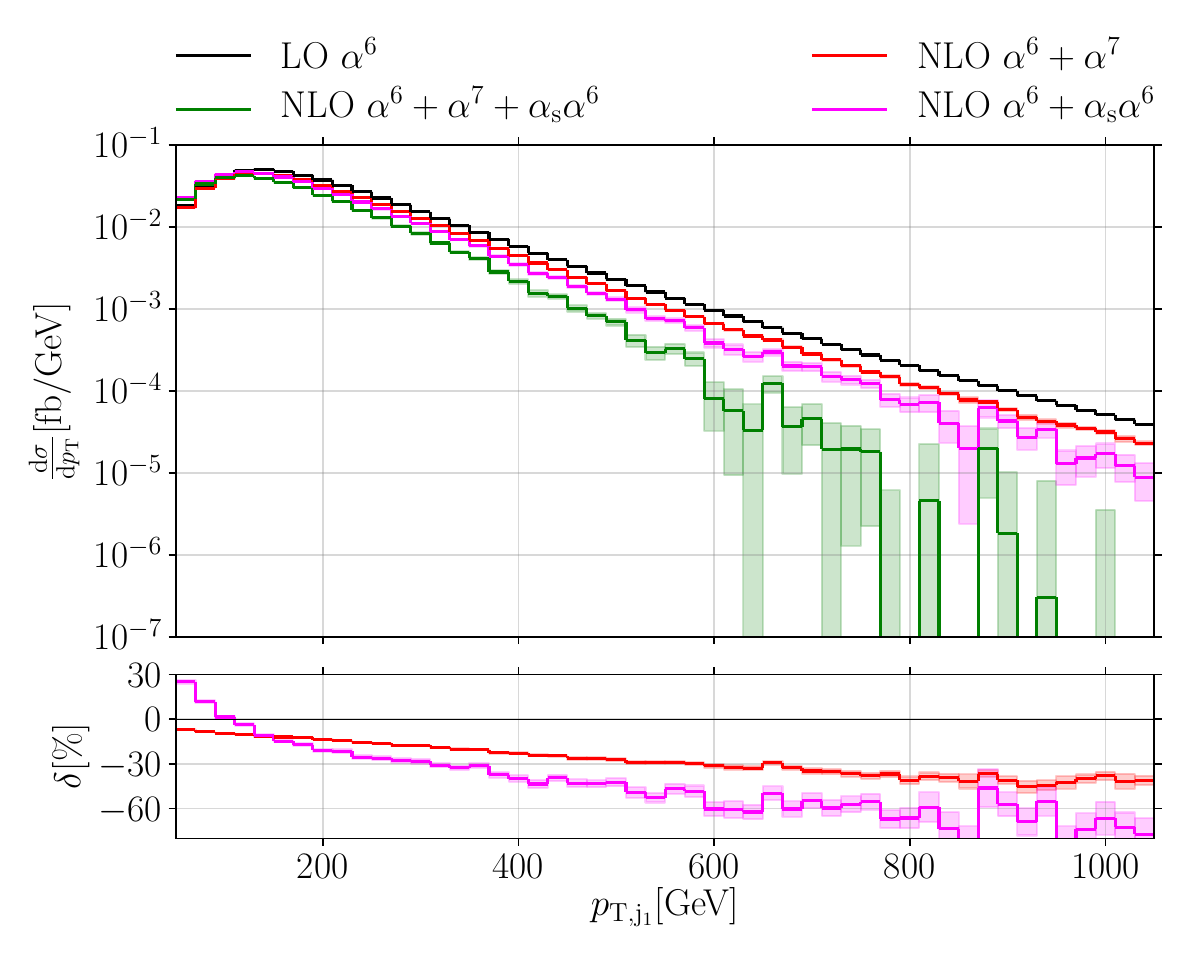}
\end{subfigure}
\hspace{\fill}
\begin{subfigure}{0.49\textwidth}
\centering
\includegraphics[page=3,width=1.\linewidth]{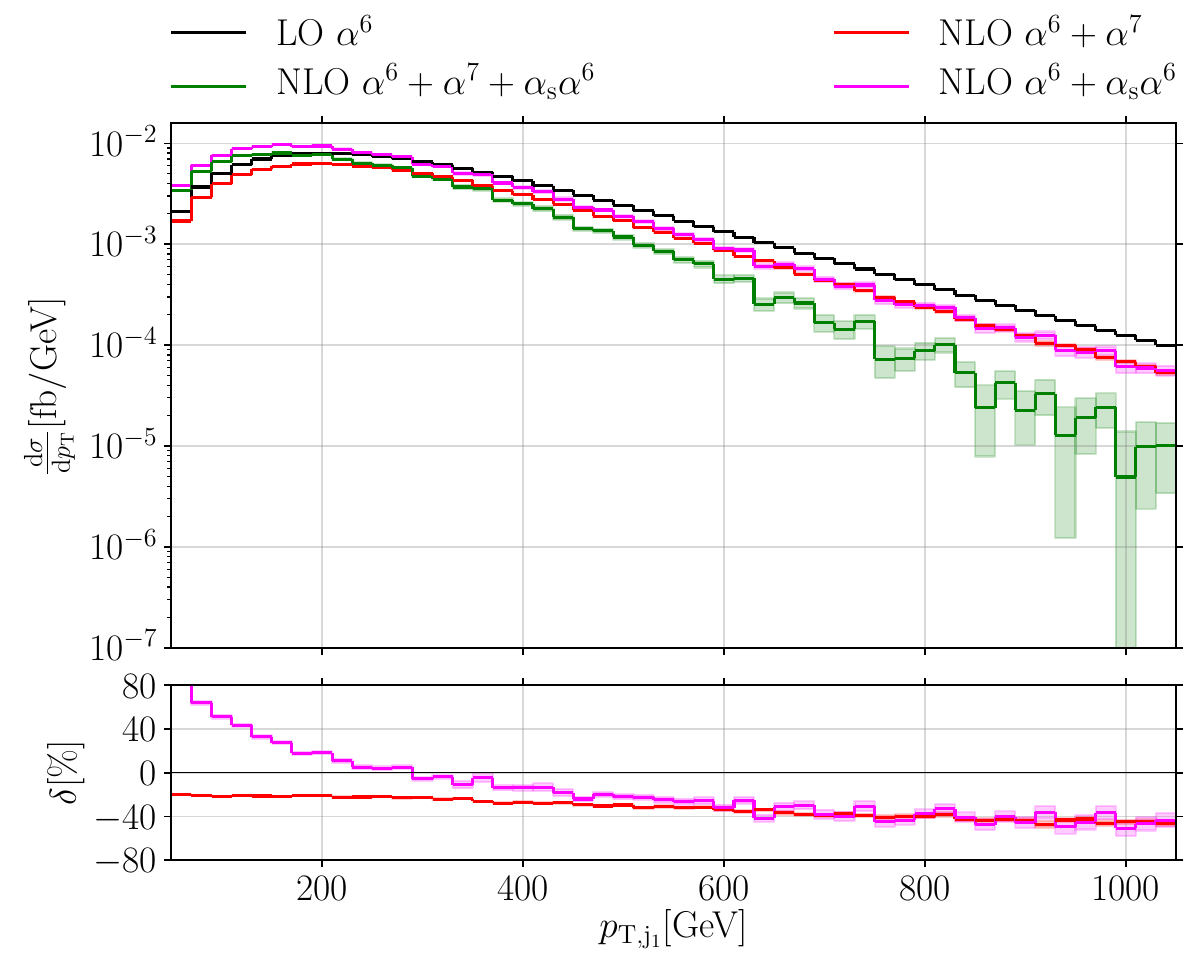}
\end{subfigure}
\caption{Differential cross section with respect to the transverse
  momentum of the leading tag jet. On the left we present the
  resolved, on the right the boosted setup. 
  Same structure as in \reffi{fig:distr_M_VV}.
} 
\label{fig:distr_pt_jtag1}
\end{figure}

In \reffi{fig:distr_pt_jtag1} we display the distribution in the
transverse momentum of the leading tag jet.
The relative NLO
corrections behave similarly in both setups. The EW corrections follow
the expected decreasing behaviour observed above in the invariant-mass
distribution of the two-vector-boson system.
The transverse momentum of the
leading jet is correlated with the energy of the VBS subprocess owing to
the $t$-channel emission of the vector bosons from the tag jets;
the harder the jet, the more energetic are the scattering vector
bosons.  Thus, the EW corrections become larger at high transverse momentum and
reach $-40\%$ in the tails at $1000\GeV$. The main difference between
the EW corrections in the two setups is at small transverse momenta,
where they are significantly larger in the boosted setup. This is due
to the fact that the boosted setup requires higher CM energies [see
Eq.~\refeq{eq:average_energy}].  The QCD corrections follow a similar
pattern: in both setups, they are positive at small transverse momenta
and become negative at $100\GeV$ in the resolved and at $300\GeV$ in the
boosted setup. The increase of the QCD corrections for small
transverse momentum of the leading tag jet has already been observed for
other VBS processes
\cite{Biedermann:2017bss,Denner:2019tmn,Denner:2020zit,Denner:2022pwc}
and is due to the suppression of the LO contribution in this
phase-space region. Since the QCD corrections are also of the order of
$-50\%$ at high transverse momenta, the NLO results
become unreliable in the tails, and NNLO effects have to be taken into
account.

\begin{figure}
\centering
\begin{subfigure}{0.49\textwidth}
\centering
\includegraphics[page=5,width=1.\linewidth]{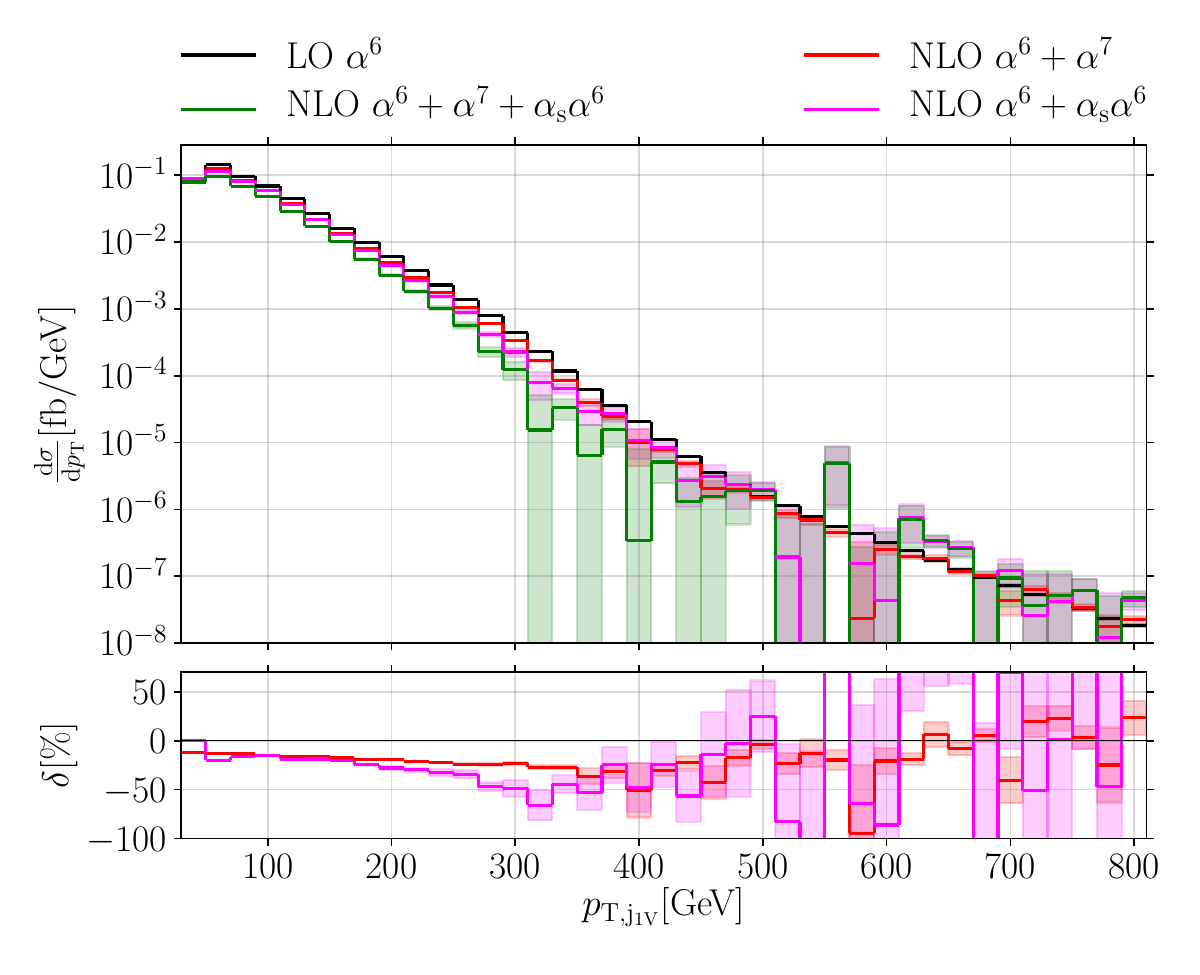}
\end{subfigure}
\hspace{\fill}
\begin{subfigure}{0.49\textwidth}
\centering
\includegraphics[page=5,width=1.\linewidth]{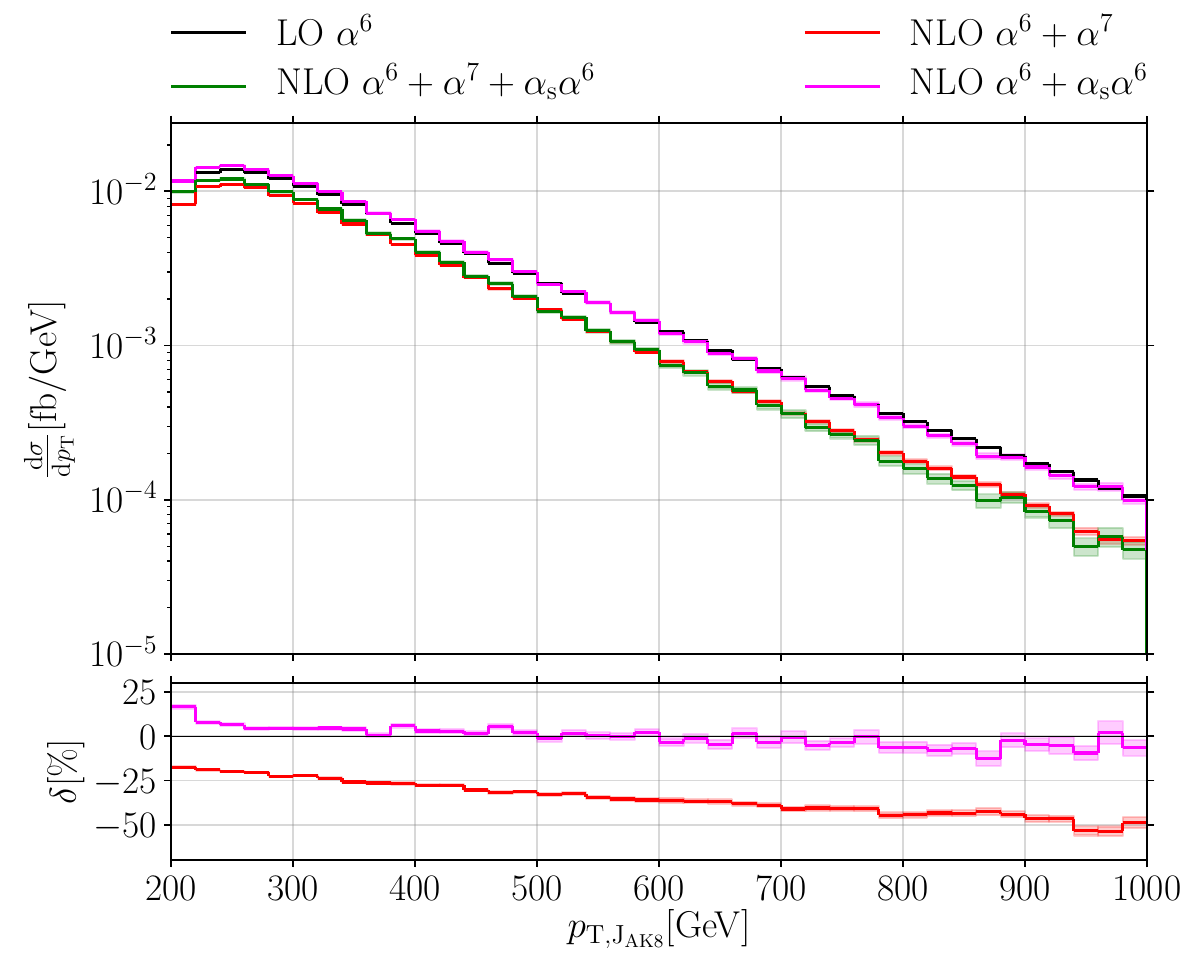}
\end{subfigure}
\caption{Differential cross section with respect to the transverse
  momentum of the leading decay jet in the resolved setup (left) and
  with respect to the transverse momentum of the AK8 jet in the
  boosted setup (right). 
  Same structure as in \reffi{fig:distr_M_VV}.
} 
\label{fig:distr_pt_jbos1}
\end{figure}

Next, we present in \reffi{fig:distr_pt_jbos1} the transverse-momentum
distribution of the leading decay jet in the resolved setup and of the
AK8 jet in the boosted setup. 
The relative NLO corrections to the two
observables exhibit a significantly different behaviour.  While for
the AK8 jet in the boosted setup the negative EW corrections increase
steadily with the transverse momentum reaching $-50\%$ in the tail and
the QCD corrections decrease monotonically, the situation is different
for the leading decay jet in the resolved setup. It was already
observed at LO in \citere{Denner:2024xul} that the differential cross section with respect to the
transverse momentum of the AK8 jet falls with an approximately
constant slope, whereas the slope of the one of the leading AK4 decay
jet changes at approximately $430\GeV$.  As a consequence of the cuts,
doubly-resonant contributions are suppressed for
$p_{\rT,\Pj_1}\gsim433\GeV$ and non-doubly resonant contributions take
over.  Consequently, below $430\GeV$, the EW
corrections show the typical behaviour for VBS processes and increase
in size
like in the boosted case up to values of $-30\%$. The QCD corrections
show a similar pattern. Above $430\GeV$, the behaviour of the
corrections completely changes as the dominant LO contribution is
different.
Obviously, the integration errors drastically increase above
$430\GeV$. This is a direct consequence and drawback of the
multichannel Monte Carlo integration, which preferably samples events
in the resonant signal regions.  However, without reading too
much into the fluctuating data, a general trend is visible: both the QCD
and the EW corrections become smaller.

\begin{figure}
\centering
\begin{subfigure}{0.49\textwidth}
\centering
\includegraphics[page=16,width=1.\linewidth]{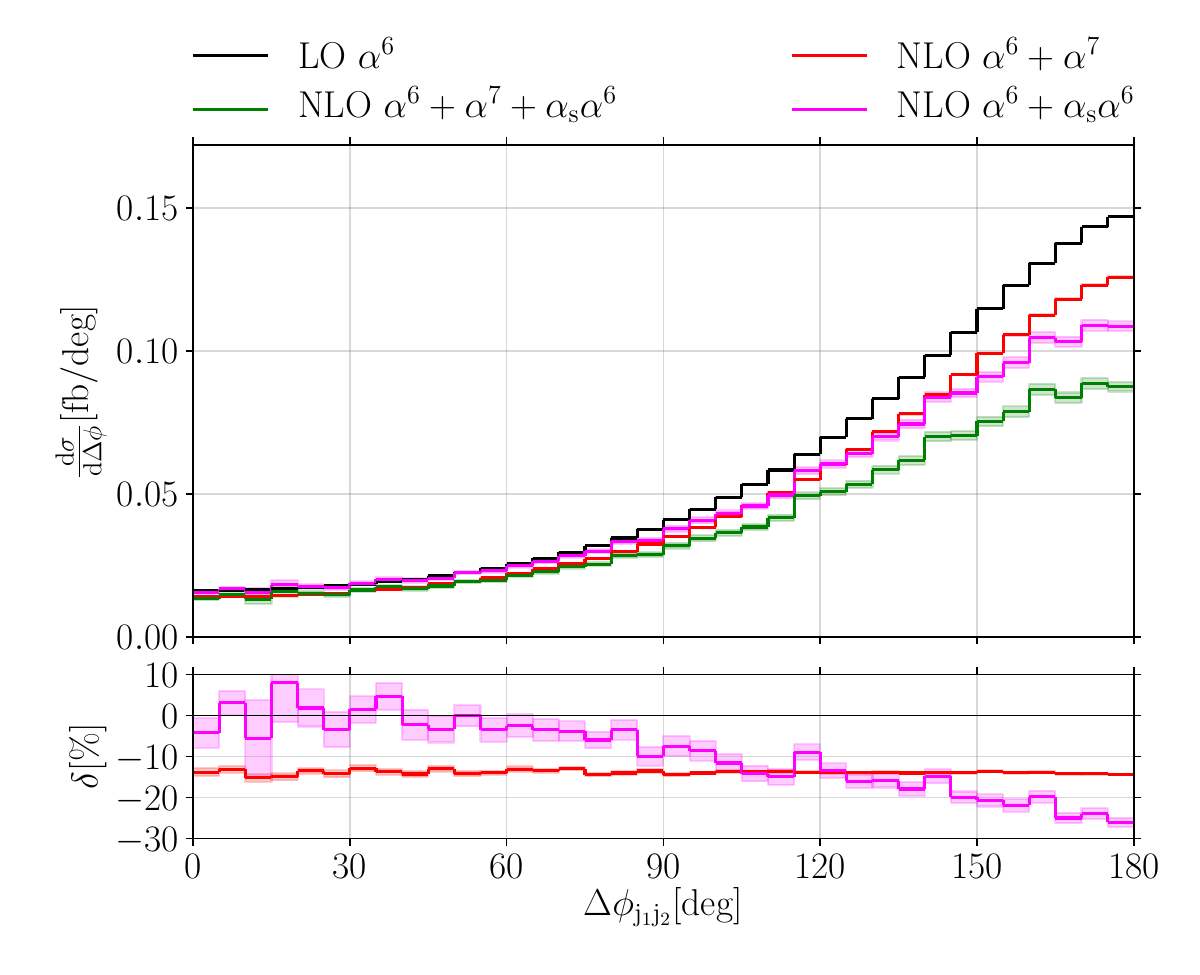}
\end{subfigure}
\hspace{\fill}
\begin{subfigure}{0.49\textwidth}
\centering
\includegraphics[page=15,width=1.\linewidth]{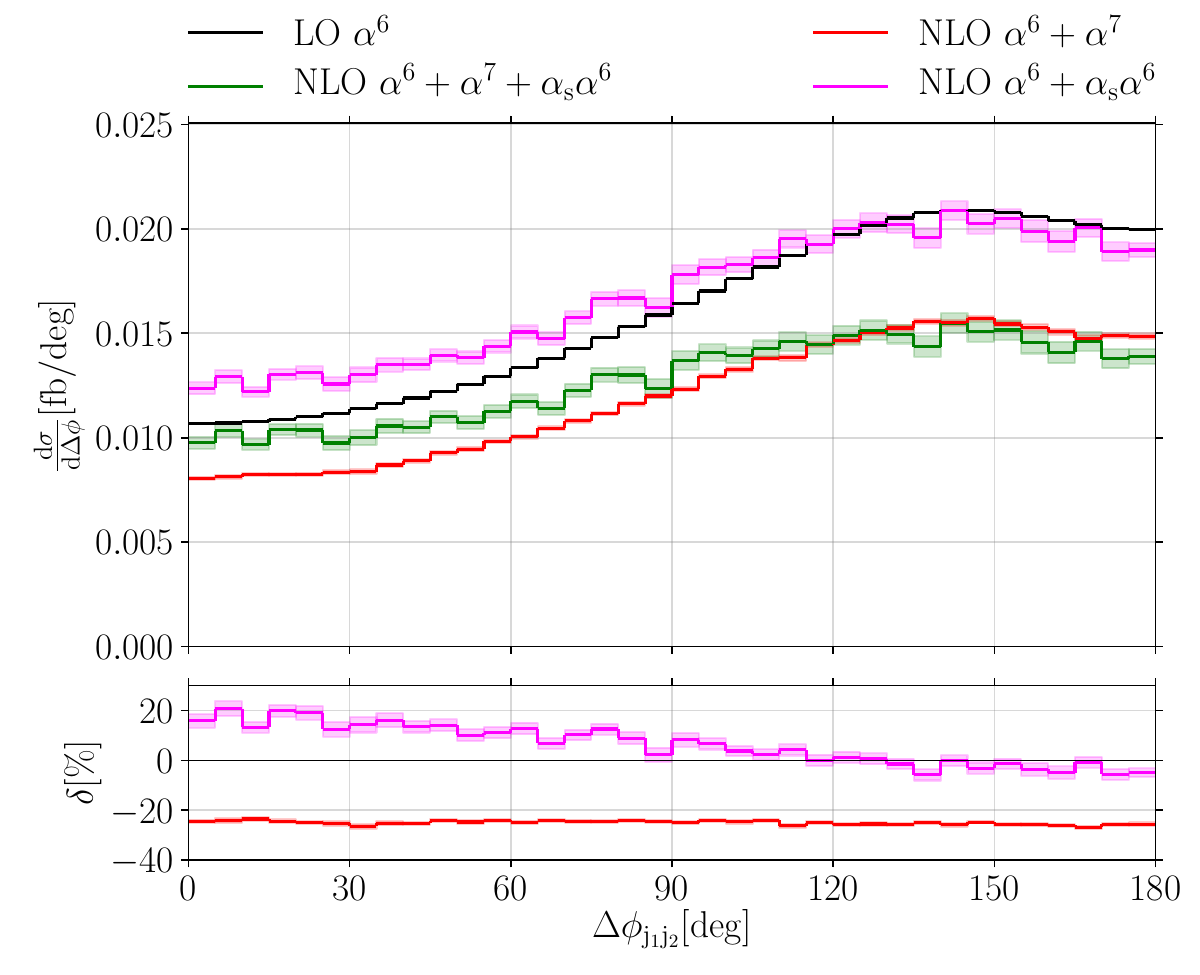}
\end{subfigure}
\caption{Differential cross section with respect to the azimuthal-angle separation of the two tag jets.   Same structure as in \reffi{fig:distr_M_VV}.}
\label{fig:distr_dphi_jj}
\end{figure}

We turn to energy-independent observables starting with the
distribution in the azimuthal-angle separation of the two tag jets in
\reffi{fig:distr_dphi_jj}. 
In both setups, the EW corrections are
completely flat over the whole phase space.  However, the relative QCD
corrections strongly depend on this observable with a similar trend in
both setups. They are more positive where the LO distribution is suppressed.
In the resolved setup, the corrections are moderate at small angular separations
and become increasingly negative at larger values, reaching $-25\%$ at
$180^\circ$. In the boosted setup, the largest positive corrections
with $20\%$ are seen at small angular separation, decreasing towards
zero at large separation.

\begin{figure}
\centering
\begin{subfigure}{0.49\textwidth}
\centering
\includegraphics[page=8,width=1.\linewidth]{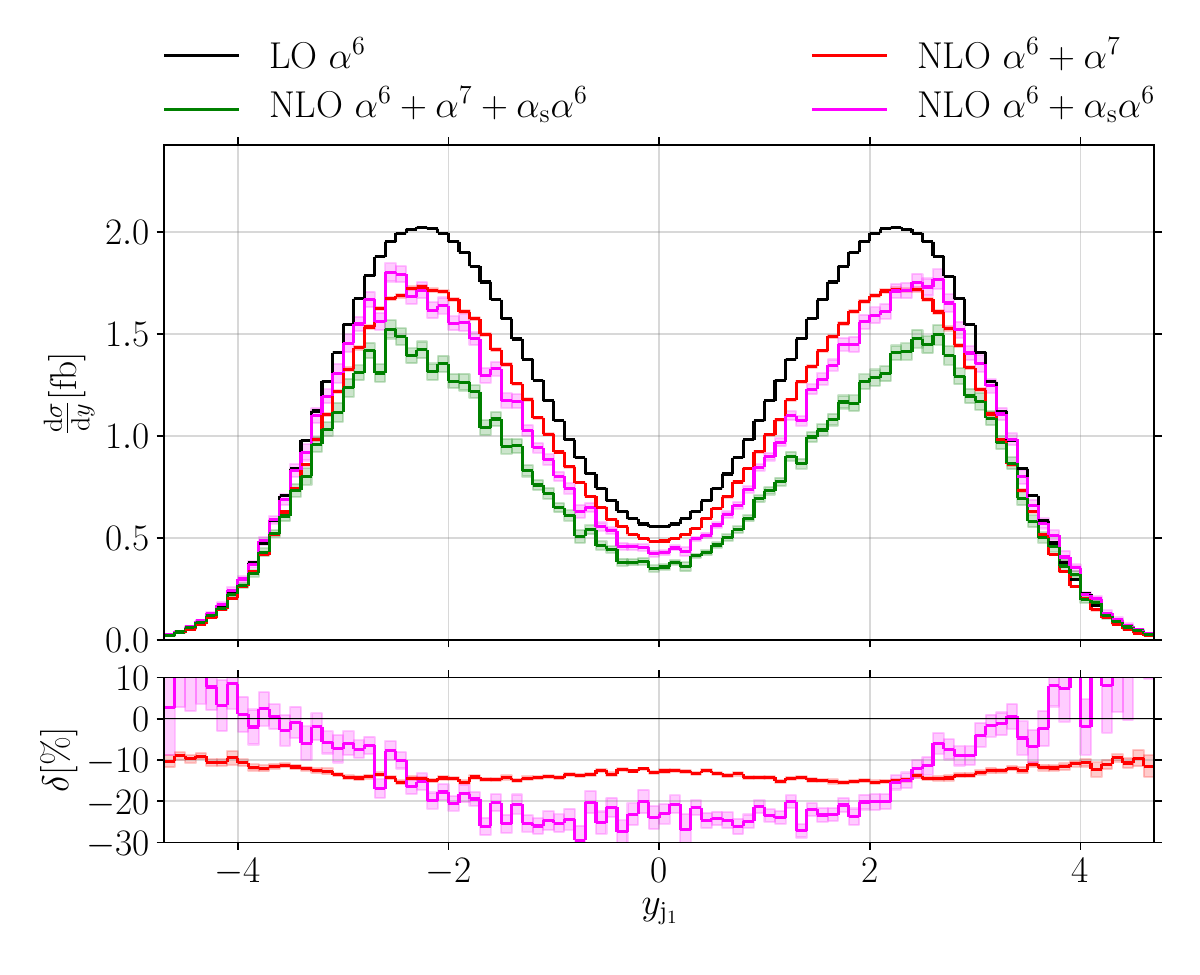}
\end{subfigure}
\hspace{\fill}
\begin{subfigure}{0.49\textwidth}
\centering
\includegraphics[page=7,width=1.\linewidth]{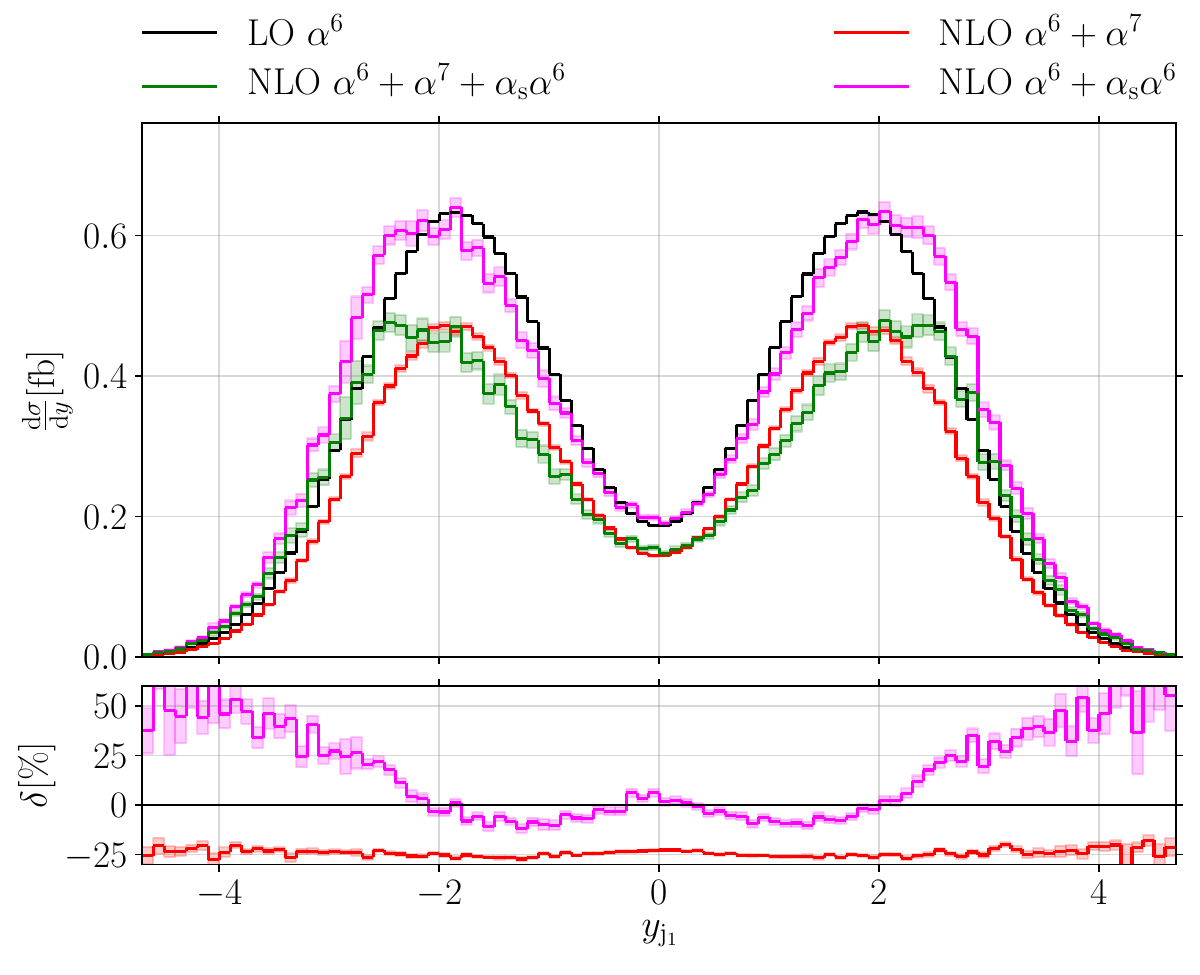}
\end{subfigure}
\caption{Differential cross section with respect to the rapidity of
  the leading tag jet. 
  Same structure as in \reffi{fig:distr_M_VV}.
}
\label{fig:distr_y_j1}
\end{figure}

Next, we discuss the distribution in the rapidity of the leading
tag jet in \reffi{fig:distr_y_j1}. 
The EW corrections show some
variation with larger negative values where the LO is large and
smaller values where it is suppressed. These variations of about $5\%$
for both the resolved and boosted setup are in accordance
with results for VBS into fully leptonic final states
\cite{Biedermann:2017bss,Denner:2019tmn,Denner:2020zit,Denner:2022pwc}.
We note that this is the energy-independent observable with the largest
variation of the relative EW corrections that we encountered.
The QCD corrections vary over a range of more than $30\%$ and $50\%$
in the resolved and boosted setup, respectively, which is larger than in fully leptonic VBS.

\begin{figure}
\centering
\begin{subfigure}{0.49\textwidth}
\centering
\includegraphics[page=18,width=1.\linewidth]{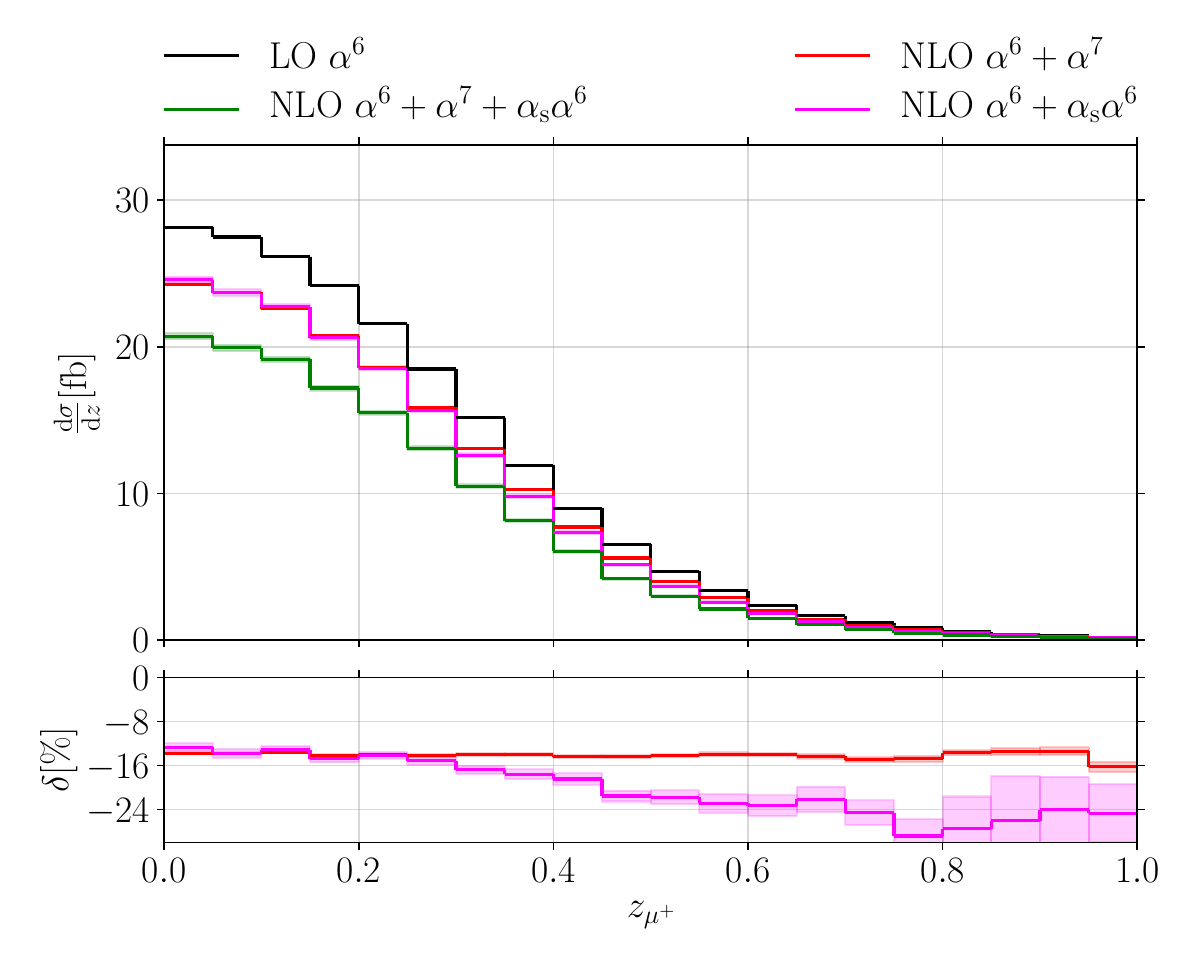}
\end{subfigure}
\hspace{\fill}
\begin{subfigure}{0.49\textwidth}
\centering
\includegraphics[page=17,width=1.\linewidth]{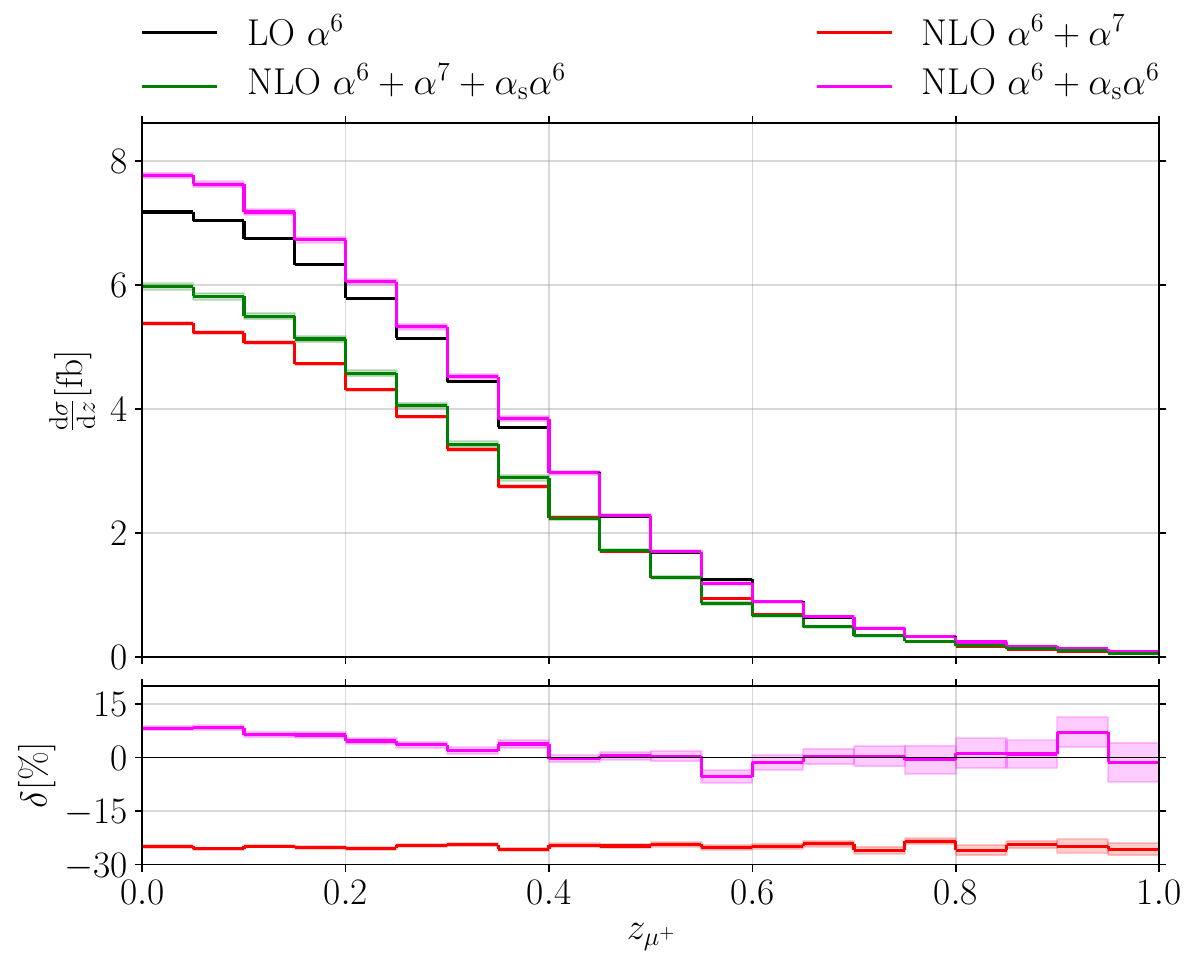}
\end{subfigure}
\caption{Differential cross section with respect to the centrality of
  the antimuon. 
  Same structure as in \reffi{fig:distr_M_VV}.
}
\label{fig:distr_z_mu}
\end{figure}

The distribution in the centrality of the antimuon with respect to the two tag jets, defined as 
\begin{equation}
z_{\mu^+} = \left|\frac{y_{\mu^+} - \frac{y_{\Pj_1} + y_{\Pj_2}}{2}}{y_{\Pj_1} - y_{\Pj_2}}\right|,
\end{equation}
is presented in \reffi{fig:distr_z_mu}.
Like in the case of the
azimuthal-angle separation, the relative EW corrections are flat over
the complete region of the phase space. The QCD corrections in the resolved
setup are always negative and become larger with higher centrality. 
In the boosted setup, the QCD corrections decrease from positive values
at small $z_{\mu^+}$ 
to zero until $z_{\mu^+} \approx 0.5$.
We note that high values of the centrality are correlated with small
values of the rapidity of the leading tag jet (see
\reffi{fig:distr_y_j1}), where the QCD corrections are also large and
negative.

\section{Conclusions}
\label{conclusions}

The scattering of vector bosons decaying into semileptonic final
states at the LHC is gaining more and more interest. In this work, we
have reported on a calculation of the NLO corrections of 
$\order{\alpha^7}$ and $\order{\alphas\alpha^6}$ to the LO EW
process of $\order{\alpha^6}$.
This has been performed employing the multichannel Monte Carlo integrator \mocanlo{},
whose results have been carefully validated against the independent in-house code \bbmc.
While the calculation is based on
the quark-induced partonic channels, we corroborated that photon-induced
channels are negligible in practice. For the included partonic
channels and perturbative orders, the calculation employs full LO and
NLO matrix elements, thus accounting for all off-shell, non-resonance,
and interference effects.
For such a final state, this is the first calculation of NLO EW
corrections, and the first one including NLO QCD corrections without any approximation. 
One of the challenges that we encountered
was to handle the large number of contributing partonic channels, which is larger than for 
the most complicated VBS processes with fully leptonic final states,
\ie $\PZ\PZ$ and $\PW^+\PW^-$ scattering, by more 
than a factor three.
We have considered two setups motivated by experimental analyses. 
In the resolved setup, the hadronically decaying vector boson is reconstructed 
from two individual decay jets, in the boosted one as a fat jet.

The EW corrections for an LHC energy of $13.6\TeV$ are found to be
large and negative, in agreement with existing results for VBS into
fully leptonic final states. The corrections to the fiducial cross section
amount to $-14\%$ in the resolved and $-25\%$ in the boosted setup,
respectively. They can be reproduced within a few percent with a
universal leading-logarithmic approximation. The larger corrections in the
boosted setup result from an event selection that enforces larger energies
for the scattering bosons.  The corrections of order
$\order{\alphas\alpha^6}$, which are predominantly but not exclusively
QCD corrections, are $-15\%$ and $+5\%$ in the resolved and boosted setup, respectively.

We have investigated the impact of the NLO corrections on various
distributions. The results generally follow the known behaviour from VBS
into fully leptonic final states. The EW corrections to distributions in
transverse momenta or invariant masses grow logarithmically with the
energy-like variable and reach values of $-40\%$ in the tails of
distributions. The variation of the EW corrections for
energy-independent observables, like angles or rapidities, remains at
the low percent level, at least in phase-space regions where the
cross sections are sizeable.

At the differential level, the QCD corrections typically range between $5\%$ and $-25\%$ in the resolved
setup and between $30\%$ and  $-5\%$ in the boosted setup, both for
energy-dependent and energy-independent observables, with even larger
contributions in some phase-space regions. In particular, for the
distribution in the invariant mass of the hadronically-decaying vector
boson, huge effects owing to the radiative return are found.
In general, the behaviour of the QCD corrections is in line with results for 
other LHC processes.

With this work, we have shown that full NLO EW and QCD calculations are
feasible for processes as complicated as VBS into semileptonic final states.
We hope that our results will be useful for the upcoming measurements
at the LHC.

\acknowledgments

We thank Sandro Uccirati and Jean-Nicolas Lang for the maintenance of
\recola and Mathieu Pellen and Giovanni Pelliccioli for useful discussions. 

We acknowledge financial support by the German Federal Ministry for Education and Research (BMBF) under contract no.\ 05H24WWA and the German Research Foundation (DFG) under reference number DE 623/8-1.
The research of DL has been supported by
the Italian Ministry of Universities and Research (MUR) under the FIS grant (CUP: D53C24005480001, FLAME).

\bibliographystyle{JHEPmod}
\bibliography{vbs_semileptonic}{}

\providecommand{\href}[2]{#2}\begingroup\raggedright\begin{thebibliography}{10}

\bibitem{ATLAS:2019cbr}
{\scshape ATLAS} collaboration, \emph{{Observation of electroweak production of
  a same-sign $W$ boson pair in association with two jets in $pp$ collisions at
  $\sqrt{s}={}$13\,TeV with the ATLAS detector}},
  \href{https://doi.org/10.1103/PhysRevLett.123.161801}{\emph{Phys. Rev. Lett.}
  {\bfseries 123} (2019) 161801}
  [\href{https://arxiv.org/abs/1906.03203}{{\ttfamily 1906.03203}}].

\bibitem{ATLAS:2023sua}
{\scshape ATLAS} collaboration, \emph{{Measurement and interpretation of
  same-sign W boson pair production in association with two jets in pp
  collisions at $ \sqrt{s}={}$13\,TeV with the ATLAS detector}},
  \href{https://doi.org/10.1007/JHEP04(2024)026}{\emph{JHEP} {\bfseries 04}
  (2024) 026} [\href{https://arxiv.org/abs/2312.00420}{{\ttfamily
  2312.00420}}].

\bibitem{CMS:2026gqm}
{\scshape CMS} collaboration, \emph{{First measurements of vector boson
  scattering in $W^\pm W^{\pm}$ and WZ production in all-leptonic final states
  at $\sqrt{s}$ = 13.6 TeV}},
  \href{https://arxiv.org/abs/2605.15396}{{\ttfamily 2605.15396}}.

\bibitem{CMS:2020gfh}
{\scshape CMS} collaboration, \emph{{Measurements of production cross sections
  of WZ and same-sign WW boson pairs in association with two jets in
  proton-proton collisions at $\sqrt{s}={}$13\,TeV}},
  \href{https://doi.org/10.1016/j.physletb.2020.135710}{\emph{Phys. Lett. B}
  {\bfseries 809} (2020) 135710}
  [\href{https://arxiv.org/abs/2005.01173}{{\ttfamily 2005.01173}}].

\bibitem{ATLAS:2024ini}
{\scshape ATLAS} collaboration, \emph{{Measurements of electroweak $W^{\pm}Z$
  boson pair production in association with two jets in $pp$ collisions at
  $\sqrt{s} ={}$ 13\,TeV with the ATLAS detector}},
  \href{https://doi.org/10.1007/JHEP06(2024)192}{\emph{JHEP} {\bfseries 06}
  (2024) 192} [\href{https://arxiv.org/abs/2403.15296}{{\ttfamily
  2403.15296}}].

\bibitem{CMS:2020fqz}
{\scshape CMS} collaboration, \emph{{Evidence for electroweak production of
  four charged leptons and two jets in proton-proton collisions at $\sqrt
  {s}={}$13\,TeV}},
  \href{https://doi.org/10.1016/j.physletb.2020.135992}{\emph{Phys. Lett. B}
  {\bfseries 812} (2021) 135992}
  [\href{https://arxiv.org/abs/2008.07013}{{\ttfamily 2008.07013}}].

\bibitem{ATLAS:2023dkz}
{\scshape ATLAS} collaboration, \emph{{Differential cross-section measurements
  of the production of four charged leptons in association with two jets using
  the ATLAS detector}},
  \href{https://doi.org/10.1007/JHEP01(2024)004}{\emph{JHEP} {\bfseries 01}
  (2024) 004} [\href{https://arxiv.org/abs/2308.12324}{{\ttfamily
  2308.12324}}].

\bibitem{CMS:2022yrl}
{\scshape CMS} collaboration, \emph{{Measurement of the electroweak production
  of W$\gamma$ in association with two jets in proton-proton collisions at
  $\sqrt{s}={}$13\,TeV}},
  \href{https://doi.org/10.1103/PhysRevD.108.032017}{\emph{Phys. Rev. D}
  {\bfseries 108} (2023) 032017}
  [\href{https://arxiv.org/abs/2212.12592}{{\ttfamily 2212.12592}}].

\bibitem{ATLAS:2024bho}
{\scshape ATLAS} collaboration, \emph{{Fiducial and differential cross-section
  measurements of electroweak $W\gamma jj$ production in $pp$ collisions at
  $\sqrt{s} ={}$ 13\,TeV with the ATLAS detector}},
  \href{https://doi.org/10.1140/epjc/s10052-024-13311-6}{\emph{Eur. Phys. J. C}
  {\bfseries 84} (2024) 1064}
  [\href{https://arxiv.org/abs/2403.02809}{{\ttfamily 2403.02809}}].

\bibitem{CMS:2021gme}
{\scshape CMS} collaboration, \emph{{Measurement of the electroweak production
  of Z$\gamma$ and two jets in proton-proton collisions at $\sqrt{s}
  ={}$13\,TeV and constraints on anomalous quartic gauge couplings}},
  \href{https://doi.org/10.1103/PhysRevD.104.072001}{\emph{Phys. Rev. D}
  {\bfseries 104} (2021) 072001}
  [\href{https://arxiv.org/abs/2106.11082}{{\ttfamily 2106.11082}}].

\bibitem{ATLAS:2023fxh}
{\scshape ATLAS} collaboration, \emph{{Measurement of the cross-sections of the
  electroweak and total production of a $Z \gamma$ pair in association with two
  jets in $pp$ collisions at $\sqrt{s}$ = 13 TeV with the ATLAS detector}},
  \href{https://doi.org/10.1016/j.physletb.2023.138222}{\emph{Phys. Lett. B}
  {\bfseries 846} (2023) 138222}
  [\href{https://arxiv.org/abs/2305.19142}{{\ttfamily 2305.19142}}].

\bibitem{CMS:2022woe}
{\scshape CMS} collaboration, \emph{{Observation of electroweak $W^+W^-$ pair
  production in association with two jets in proton-proton collisions at
  $\sqrt{s}={}$13\,TeV}},
  \href{https://doi.org/10.1016/j.physletb.2022.137495}{\emph{Phys. Lett. B}
  {\bfseries 841} (2023) 137495}
  [\href{https://arxiv.org/abs/2205.05711}{{\ttfamily 2205.05711}}].

\bibitem{ATLAS:2024ett}
{\scshape ATLAS} collaboration, \emph{{Observation of electroweak production of
  $W^+W^-$ in association with jets in proton-proton collisions at
  $\sqrt{s}={}$13\,TeV with the ATLAS Detector}},
  \href{https://doi.org/10.1007/JHEP07(2024)254}{\emph{JHEP} {\bfseries 07}
  (2024) 254} [\href{https://arxiv.org/abs/2403.04869}{{\ttfamily
  2403.04869}}].

\bibitem{CMS:2020etf}
{\scshape CMS} collaboration, \emph{{Measurements of production cross sections
  of polarized same-sign W boson pairs in association with two jets in
  proton-proton collisions at $\sqrt{s} ={}$13\,TeV}},
  \href{https://doi.org/10.1016/j.physletb.2020.136018}{\emph{Phys. Lett. B}
  {\bfseries 812} (2021) 136018}
  [\href{https://arxiv.org/abs/2009.09429}{{\ttfamily 2009.09429}}].

\bibitem{ATLAS:2025wuw}
{\scshape ATLAS} collaboration, \emph{{Evidence for Longitudinally Polarized W
  Bosons in the Electroweak Production of Same-Sign W Boson Pairs in
  Association with Two Jets in pp Collisions at s=13{\,}{\,}TeV with the ATLAS
  Detector}}, \href{https://doi.org/10.1103/bpln-ccql}{\emph{Phys. Rev. Lett.}
  {\bfseries 135} (2025) 111802}
  [\href{https://arxiv.org/abs/2503.11317}{{\ttfamily 2503.11317}}].

\bibitem{Aaboud:2019gxl}
{\scshape ATLAS} collaboration, \emph{{Measurement of $W^{\pm}Z$ production
  cross sections and gauge boson polarisation in $pp$ collisions at $\sqrt{s}
  ={}$13\,TeV with the ATLAS detector}},
  \href{https://doi.org/10.1140/epjc/s10052-019-7027-6}{\emph{Eur. Phys. J. C}
  {\bfseries 79} (2019) 535}
  [\href{https://arxiv.org/abs/1902.05759}{{\ttfamily 1902.05759}}].

\bibitem{CMS:2021icx}
{\scshape CMS} collaboration, \emph{{Measurement of the inclusive and
  differential WZ production cross sections, polarization angles, and triple
  gauge couplings in pp collisions at $ \sqrt{s}=$13\,TeV}},
  \href{https://doi.org/10.1007/JHEP07(2022)032}{\emph{JHEP} {\bfseries 07}
  (2022) 032} [\href{https://arxiv.org/abs/2110.11231}{{\ttfamily
  2110.11231}}].

\bibitem{ATLAS:2022oge}
{\scshape ATLAS} collaboration, \emph{{Observation of gauge boson
  joint-polarisation states in $W^{\pm}Z$ production from $pp$ collisions at
  $\sqrt{s} = 13$ TeV with the ATLAS detector}},
  \href{https://doi.org/10.1016/j.physletb.2023.137895}{\emph{Phys. Lett. B}
  {\bfseries 843} (2023) 137895}
  [\href{https://arxiv.org/abs/2211.09435}{{\ttfamily 2211.09435}}].

\bibitem{ATLAS:2024qbd}
{\scshape ATLAS} collaboration, \emph{{Studies of the Energy Dependence of
  Diboson Polarization Fractions and the Radiation-Amplitude-Zero Effect in WZ
  Production with the ATLAS Detector}},
  \href{https://doi.org/10.1103/PhysRevLett.133.101802}{\emph{Phys. Rev. Lett.}
  {\bfseries 133} (2024) 101802}
  [\href{https://arxiv.org/abs/2402.16365}{{\ttfamily 2402.16365}}].

\bibitem{ATLAS:2023zrv}
{\scshape ATLAS} collaboration, \emph{{Evidence of pair production of
  longitudinally polarised vector bosons and study of CP properties in ZZ
  $\rightarrow{} 4\ensuremath{\ell}$ events with the ATLAS detector at $
  \sqrt{s} $ = 13 TeV}},
  \href{https://doi.org/10.1007/JHEP12(2023)107}{\emph{JHEP} {\bfseries 12}
  (2023) 107} [\href{https://arxiv.org/abs/2310.04350}{{\ttfamily
  2310.04350}}].

\bibitem{CMS:2019qfk}
{\scshape CMS} collaboration, \emph{{Search for anomalous electroweak
  production of vector boson pairs in association with two jets in
  proton-proton collisions at 13\,TeV}},
  \href{https://doi.org/10.1016/j.physletb.2019.134985}{\emph{Phys. Lett. B}
  {\bfseries 798} (2019) 134985}
  [\href{https://arxiv.org/abs/1905.07445}{{\ttfamily 1905.07445}}].

\bibitem{ATLAS:2019thr}
{\scshape ATLAS} collaboration, \emph{{Search for the electroweak diboson
  production in association with a high-mass dijet system in semileptonic final
  states in $pp$ collisions at $\sqrt{s}={}$13\,TeV with the ATLAS detector}},
  \href{https://doi.org/10.1103/PhysRevD.100.032007}{\emph{Phys. Rev. D}
  {\bfseries 100} (2019) 032007}
  [\href{https://arxiv.org/abs/1905.07714}{{\ttfamily 1905.07714}}].

\bibitem{CMS:2021qzzv2}
{\scshape CMS} collaboration, \emph{{Evidence for WW/WZ vector boson scattering
  in the decay channel \ensuremath{\ell}\ensuremath{\nu}qq produced in
  association with two jets in proton-proton collisions at
  $\sqrt{s}={}$13\,TeV}},
  \href{https://doi.org/10.1016/j.physletb.2022.137438}{\emph{Phys. Lett. B}
  {\bfseries 834} (2022) 137438}
  [\href{https://arxiv.org/abs/2112.05259v2}{{\ttfamily 2112.05259v2}}].

\bibitem{ATLAS:2025omi}
{\scshape ATLAS} collaboration, \emph{{Electroweak diboson production in
  association with a high-mass dijet system in semileptonic final states from
  $pp$ collisions at $\sqrt{s} = 13$ TeV with the ATLAS detector}},
  \href{https://doi.org/10.1140/epjc/s10052-025-15217-3}{\emph{Eur. Phys. J. C}
  {\bfseries 86} (2026) 433}
  [\href{https://arxiv.org/abs/2503.17461}{{\ttfamily 2503.17461}}].

\bibitem{Jager:2006zc}
B.~J{\"a}ger, C.~Oleari and D.~Zeppenfeld, \emph{{Next-to-leading order QCD
  corrections to $W^+W^-$ production via vector-boson fusion}},
  \href{https://doi.org/10.1088/1126-6708/2006/07/015}{\emph{JHEP} {\bfseries
  07} (2006) 015} [\href{https://arxiv.org/abs/hep-ph/0603177}{{\ttfamily
  hep-ph/0603177}}].

\bibitem{Jager:2006cp}
B.~J{\"a}ger, C.~Oleari and D.~Zeppenfeld, \emph{{Next-to-leading order QCD
  corrections to Z boson pair production via vector-boson fusion}},
  \href{https://doi.org/10.1103/PhysRevD.73.113006}{\emph{Phys. Rev. D}
  {\bfseries 73} (2006) 113006}
  [\href{https://arxiv.org/abs/hep-ph/0604200}{{\ttfamily hep-ph/0604200}}].

\bibitem{Bozzi:2007ur}
G.~Bozzi, B.~J{\"a}ger, C.~Oleari and D.~Zeppenfeld, \emph{{Next-to-leading
  order QCD corrections to $W^+ Z$ and $W^- Z$ production via vector-boson
  fusion}}, \href{https://doi.org/10.1103/PhysRevD.75.073004}{\emph{Phys. Rev.
  D} {\bfseries 75} (2007) 073004}
  [\href{https://arxiv.org/abs/hep-ph/0701105}{{\ttfamily hep-ph/0701105}}].

\bibitem{Jager:2009xx}
B.~J{\"a}ger, C.~Oleari and D.~Zeppenfeld, \emph{{Next-to-leading order QCD
  corrections to $W^+ W^+ jj$ and $W^- W^- jj$ production via weak-boson
  fusion}}, \href{https://doi.org/10.1103/PhysRevD.80.034022}{\emph{Phys. Rev.
  D} {\bfseries 80} (2009) 034022}
  [\href{https://arxiv.org/abs/0907.0580}{{\ttfamily 0907.0580}}].

\bibitem{Denner:2012dz}
A.~Denner, L.~Ho\v{s}ekov\'a and S.~Kallweit, \emph{{NLO QCD corrections to
  $W^+W^+$jj production in vector-boson fusion at the LHC}},
  \href{https://doi.org/10.1103/PhysRevD.86.114014}{\emph{Phys. Rev. D}
  {\bfseries 86} (2012) 114014}
  [\href{https://arxiv.org/abs/1209.2389}{{\ttfamily 1209.2389}}].

\bibitem{Rauch:2016pai}
M.~Rauch, \emph{{Vector-Boson Fusion and Vector-Boson Scattering}},
  \href{https://arxiv.org/abs/1610.08420}{{\ttfamily 1610.08420}}.

\bibitem{Melia:2010bm}
T.~Melia, K.~Melnikov, R.~R{\"o}ntsch and G.~Zanderighi, \emph{{Next-to-leading
  order QCD predictions for $W^+W^+jj$ production at the LHC}},
  \href{https://doi.org/10.1007/JHEP12(2010)053}{\emph{JHEP} {\bfseries 12}
  (2010) 053} [\href{https://arxiv.org/abs/1007.5313}{{\ttfamily 1007.5313}}].

\bibitem{Melia:2011dw}
T.~Melia, K.~Melnikov, R.~R{\"o}ntsch and G.~Zanderighi, \emph{{NLO QCD
  corrections for $W^+W^-$ pair production in association with two jets at
  hadron colliders}},
  \href{https://doi.org/10.1103/PhysRevD.83.114043}{\emph{Phys. Rev. D}
  {\bfseries 83} (2011) 114043}
  [\href{https://arxiv.org/abs/1104.2327}{{\ttfamily 1104.2327}}].

\bibitem{Greiner:2012im}
N.~Greiner, G.~Heinrich, P.~Mastrolia, G.~Ossola, T.~Reiter and F.~Tramontano,
  \emph{{NLO QCD corrections to the production of $W^+ W^-$ plus two jets at
  the LHC}}, \href{https://doi.org/10.1016/j.physletb.2012.06.027}{\emph{Phys.
  Lett. B} {\bfseries 713} (2012) 277}
  [\href{https://arxiv.org/abs/1202.6004}{{\ttfamily 1202.6004}}].

\bibitem{Campanario:2013qba}
F.~Campanario, M.~Kerner, L.D.~Ninh and D.~Zeppenfeld, \emph{{WZ Production in
  Association with Two Jets at Next-to-Leading Order in QCD}},
  \href{https://doi.org/10.1103/PhysRevLett.111.052003}{\emph{Phys. Rev. Lett.}
  {\bfseries 111} (2013) 052003}
  [\href{https://arxiv.org/abs/1305.1623}{{\ttfamily 1305.1623}}].

\bibitem{Campanario:2014ioa}
F.~Campanario, M.~Kerner, L.D.~Ninh and D.~Zeppenfeld, \emph{{Next-to-leading
  order QCD corrections to ZZ production in association with two jets}},
  \href{https://doi.org/10.1007/JHEP07(2014)148}{\emph{JHEP} {\bfseries 07}
  (2014) 148} [\href{https://arxiv.org/abs/1405.3972}{{\ttfamily 1405.3972}}].

\bibitem{Campanario:2014dpa}
F.~Campanario, M.~Kerner, L.D.~Ninh and D.~Zeppenfeld, \emph{{Next-to-leading
  order QCD corrections to $W \gamma$ production in association with two
  jets}}, \href{https://doi.org/10.1140/epjc/s10052-014-2882-7}{\emph{Eur.
  Phys. J. C} {\bfseries 74} (2014) 2882}
  [\href{https://arxiv.org/abs/1402.0505}{{\ttfamily 1402.0505}}].

\bibitem{Ballestrero:2018anz}
A.~Ballestrero et~al., \emph{{Precise predictions for same-sign W-boson
  scattering at the LHC}},
  \href{https://doi.org/10.1140/epjc/s10052-018-6136-y}{\emph{Eur. Phys. J. C}
  {\bfseries 78} (2018) 671}
  [\href{https://arxiv.org/abs/1803.07943}{{\ttfamily 1803.07943}}].

\bibitem{Jager:2011ms}
B.~J{\"a}ger and G.~Zanderighi, \emph{{NLO corrections to electroweak and QCD
  production of $W^+W^+$ plus two jets in the POWHEGBOX}},
  \href{https://doi.org/10.1007/JHEP11(2011)055}{\emph{JHEP} {\bfseries 11}
  (2011) 055} [\href{https://arxiv.org/abs/1108.0864}{{\ttfamily 1108.0864}}].

\bibitem{Jager:2013mu}
B.~J{\"a}ger and G.~Zanderighi, \emph{{Electroweak $W^+W^-jj$ prodution at NLO
  in QCD matched with parton shower in the POWHEG-BOX}},
  \href{https://doi.org/10.1007/JHEP04(2013)024}{\emph{JHEP} {\bfseries 04}
  (2013) 024} [\href{https://arxiv.org/abs/1301.1695}{{\ttfamily 1301.1695}}].

\bibitem{Jager:2013iza}
B.~J\"ager, A.~Karlberg and G.~Zanderighi, \emph{{Electroweak $ZZjj$ production
  in the Standard Model and beyond in the POWHEG-BOX V2}},
  \href{https://doi.org/10.1007/JHEP03(2014)141}{\emph{JHEP} {\bfseries 03}
  (2014) 141} [\href{https://arxiv.org/abs/1312.3252}{{\ttfamily 1312.3252}}].

\bibitem{Rauch:2016upa}
M.~Rauch and S.~Pl\"atzer, \emph{{Parton Shower Matching Systematics in
  Vector-Boson-Fusion WW Production}},
  \href{https://doi.org/10.1140/epjc/s10052-017-4860-3}{\emph{Eur. Phys. J. C}
  {\bfseries 77} (2017) 293}
  [\href{https://arxiv.org/abs/1605.07851}{{\ttfamily 1605.07851}}].

\bibitem{Rauch:2016jxo}
M.~Rauch and S.~Pl\"atzer, \emph{{Parton-shower Effects in Vector-Boson-Fusion
  Processes}}, \href{https://doi.org/10.22323/1.265.0076}{\emph{PoS} {\bfseries
  DIS2016} (2016) 076} [\href{https://arxiv.org/abs/1607.00159}{{\ttfamily
  1607.00159}}].

\bibitem{Jager:2018cyo}
B.~J{\"a}ger, A.~Karlberg and J.~Scheller, \emph{{Parton-shower effects in
  electroweak $WZjj$ production at the next-to-leading order of QCD}},
  \href{https://doi.org/10.1140/epjc/s10052-019-6736-1}{\emph{Eur. Phys. J. C}
  {\bfseries 79} (2019) 226}
  [\href{https://arxiv.org/abs/1812.05118}{{\ttfamily 1812.05118}}].

\bibitem{Jager:2024sij}
B.~J\"ager, A.~Karlberg and S.~Reinhardt, \emph{{QCD effects in electroweak
  $WZjj$ production at current and future hadron colliders}},
  \href{https://doi.org/10.1140/epjc/s10052-024-12920-5}{\emph{Eur. Phys. J. C}
  {\bfseries 84} (2024) 587}
  [\href{https://arxiv.org/abs/2403.12192}{{\ttfamily 2403.12192}}].

\bibitem{Melia:2011tj}
T.~Melia, P.~Nason, R.~R{\"o}ntsch and G.~Zanderighi, \emph{{$W^+W^-$, WZ and
  ZZ production in the POWHEG BOX}},
  \href{https://doi.org/10.1007/JHEP11(2011)078}{\emph{JHEP} {\bfseries 11}
  (2011) 078} [\href{https://arxiv.org/abs/1107.5051}{{\ttfamily 1107.5051}}].

\bibitem{Baglio:2014uba}
J.~Baglio et~al., \emph{{Release Note - VBFNLO 2.7.0}},
  \href{https://arxiv.org/abs/1404.3940}{{\ttfamily 1404.3940}}.

\bibitem{Stelzer:1994ta}
T.~Stelzer and W.F.~Long, \emph{{Automatic generation of tree level helicity
  amplitudes}},
  \href{https://doi.org/10.1016/0010-4655(94)90084-1}{\emph{Comput. Phys.
  Commun.} {\bfseries 81} (1994) 357}
  [\href{https://arxiv.org/abs/hep-ph/9401258}{{\ttfamily hep-ph/9401258}}].

\bibitem{Alwall:2014hca}
J.~Alwall, R.~Frederix, S.~Frixione, V.~Hirschi, F.~Maltoni, O.~Mattelaer
  et~al., \emph{{The automated computation of tree-level and next-to-leading
  order differential cross sections, and their matching to parton shower
  simulations}}, \href{https://doi.org/10.1007/JHEP07(2014)079}{\emph{JHEP}
  {\bfseries 07} (2014) 079} [\href{https://arxiv.org/abs/1405.0301}{{\ttfamily
  1405.0301}}].

\bibitem{Sherpa:2019gpd}
{\scshape Sherpa} collaboration, \emph{{Event Generation with Sherpa 2.2}},
  \href{https://doi.org/10.21468/SciPostPhys.7.3.034}{\emph{SciPost Phys.}
  {\bfseries 7} (2019) 034} [\href{https://arxiv.org/abs/1905.09127}{{\ttfamily
  1905.09127}}].

\bibitem{Nason:2004rx}
P.~Nason, \emph{{A new method for combining NLO QCD with shower Monte Carlo
  algorithms}},
  \href{https://doi.org/10.1088/1126-6708/2004/11/040}{\emph{JHEP} {\bfseries
  11} (2004) 040} [\href{https://arxiv.org/abs/hep-ph/0409146}{{\ttfamily
  hep-ph/0409146}}].

\bibitem{Frixione:2007vw}
S.~Frixione, P.~Nason and C.~Oleari, \emph{{Matching NLO QCD computations with
  Parton Shower simulations: the POWHEG method}},
  \href{https://doi.org/10.1088/1126-6708/2007/11/070}{\emph{JHEP} {\bfseries
  11} (2007) 070} [\href{https://arxiv.org/abs/0709.2092}{{\ttfamily
  0709.2092}}].

\bibitem{Alioli:2010xd}
S.~Alioli, P.~Nason, C.~Oleari and E.~Re, \emph{{A general framework for
  implementing NLO calculations in shower Monte Carlo programs: the POWHEG
  BOX}}, \href{https://doi.org/10.1007/JHEP06(2010)043}{\emph{JHEP} {\bfseries
  06} (2010) 043} [\href{https://arxiv.org/abs/1002.2581}{{\ttfamily
  1002.2581}}].

\bibitem{Biedermann:2016yds}
B.~Biedermann, A.~Denner and M.~Pellen, \emph{{Large electroweak corrections to
  vector-boson scattering at the Large Hadron Collider}},
  \href{https://doi.org/10.1103/PhysRevLett.118.261801}{\emph{Phys. Rev. Lett.}
  {\bfseries 118} (2017) 261801}
  [\href{https://arxiv.org/abs/1611.02951}{{\ttfamily 1611.02951}}].

\bibitem{Denner:2019tmn}
A.~Denner, S.~Dittmaier, P.~Maierh\"ofer, M.~Pellen and C.~Schwan, \emph{{QCD
  and electroweak corrections to WZ scattering at the LHC}},
  \href{https://doi.org/10.1007/JHEP06(2019)067}{\emph{JHEP} {\bfseries 06}
  (2019) 067} [\href{https://arxiv.org/abs/1904.00882}{{\ttfamily
  1904.00882}}].

\bibitem{Denner:2020zit}
A.~Denner, R.~Franken, M.~Pellen and T.~Schmidt, \emph{{NLO QCD and EW
  corrections to vector-boson scattering into ZZ at the LHC}},
  \href{https://doi.org/10.1007/JHEP11(2020)110}{\emph{JHEP} {\bfseries 11}
  (2020) 110} [\href{https://arxiv.org/abs/2009.00411}{{\ttfamily
  2009.00411}}].

\bibitem{Denner:2022pwc}
A.~Denner, R.~Franken, T.~Schmidt and C.~Schwan, \emph{{NLO QCD and EW
  corrections to vector-boson scattering into $W^{+}W^{-}$ at the LHC}},
  \href{https://doi.org/10.1007/JHEP06(2022)098}{\emph{JHEP} {\bfseries 06}
  (2022) 098} [\href{https://arxiv.org/abs/2202.10844}{{\ttfamily
  2202.10844}}].

\bibitem{Biedermann:2017bss}
B.~Biedermann, A.~Denner and M.~Pellen, \emph{{Complete NLO corrections to
  $W^{+}W^{+}$ scattering and its irreducible background at the LHC}},
  \href{https://doi.org/10.1007/JHEP10(2017)124}{\emph{JHEP} {\bfseries 10}
  (2017) 124} [\href{https://arxiv.org/abs/1708.00268}{{\ttfamily
  1708.00268}}].

\bibitem{Denner:2021hsa}
A.~Denner, R.~Franken, M.~Pellen and T.~Schmidt, \emph{{Full NLO predictions
  for vector-boson scattering into Z bosons and its irreducible background at
  the LHC}}, \href{https://doi.org/10.1007/JHEP10(2021)228}{\emph{JHEP}
  {\bfseries 10} (2021) 228}
  [\href{https://arxiv.org/abs/2107.10688}{{\ttfamily 2107.10688}}].

\bibitem{Denner:2026phn}
A.~Denner, D.~Lombardi, S.~Lopez Portillo~Chavez, M.~Pellen and G.~Pelliccioli,
  \emph{{MoCaNLO: a Monte Carlo integrator for NLO calculations}},
  \href{https://arxiv.org/abs/2602.19842}{{\ttfamily 2602.19842}}.

\bibitem{denner_2026_19829093}
A.~Denner, D.~Lombardi, S.~Lopez Portillo~Chavez, M.~Pellen and G.~Pelliccioli,
  \emph{Mocanlo},  Feb., 2026.
\newblock 10.5281/zenodo.19829093.

\bibitem{Chiesa:2019ulk}
M.~Chiesa, A.~Denner, J.-N.~Lang and M.~Pellen, \emph{{An event generator for
  same-sign W-boson scattering at the LHC including electroweak corrections}},
  \href{https://doi.org/10.1140/epjc/s10052-019-7290-6}{\emph{Eur. Phys. J. C}
  {\bfseries 79} (2019) 788}
  [\href{https://arxiv.org/abs/1906.01863}{{\ttfamily 1906.01863}}].

\bibitem{Ballestrero:2020qgv}
A.~Ballestrero, E.~Maina and G.~Pelliccioli, \emph{{Different polarization
  definitions in same-sign $WW$ scattering at the LHC}},
  \href{https://doi.org/10.1016/j.physletb.2020.135856}{\emph{Phys. Lett. B}
  {\bfseries 811} (2020) 135856}
  [\href{https://arxiv.org/abs/2007.07133}{{\ttfamily 2007.07133}}].

\bibitem{Ballestrero:2017bxn}
A.~Ballestrero, E.~Maina and G.~Pelliccioli, \emph{{$W$ boson polarization in
  vector boson scattering at the LHC}},
  \href{https://doi.org/10.1007/JHEP03(2018)170}{\emph{JHEP} {\bfseries 03}
  (2018) 170} [\href{https://arxiv.org/abs/1710.09339}{{\ttfamily
  1710.09339}}].

\bibitem{Ballestrero:2019qoy}
A.~Ballestrero, E.~Maina and G.~Pelliccioli, \emph{{Polarized vector boson
  scattering in the fully leptonic WZ and ZZ channels at the LHC}},
  \href{https://doi.org/10.1007/JHEP09(2019)087}{\emph{JHEP} {\bfseries 09}
  (2019) 087} [\href{https://arxiv.org/abs/1907.04722}{{\ttfamily
  1907.04722}}].

\bibitem{BuarqueFranzosi:2019boy}
D.~Buarque~Franzosi, O.~Mattelaer, R.~Ruiz and S.~Shil, \emph{{Automated
  predictions from polarized matrix elements}},
  \href{https://doi.org/10.1007/JHEP04(2020)082}{\emph{JHEP} {\bfseries 04}
  (2020) 082} [\href{https://arxiv.org/abs/1912.01725}{{\ttfamily
  1912.01725}}].

\bibitem{Denner:2024tlu}
A.~Denner, C.~Haitz and G.~Pelliccioli, \emph{{NLO EW and QCD corrections to
  polarised same-sign WW scattering at the LHC}},
  \href{https://doi.org/10.1007/JHEP11(2024)115}{\emph{JHEP} {\bfseries 11}
  (2024) 115} [\href{https://arxiv.org/abs/2409.03620}{{\ttfamily
  2409.03620}}].

\bibitem{Denner:2025xdz}
A.~Denner, R.~Franken, C.~Haitz, D.~Lombardi and G.~Pelliccioli,
  \emph{{Electroweak corrections to doubly polarised WZ scattering at the
  LHC}}, \href{https://doi.org/10.1007/JHEP02(2026)120}{\emph{JHEP} {\bfseries
  02} (2026) 120} [\href{https://arxiv.org/abs/2510.26462}{{\ttfamily
  2510.26462}}].

\bibitem{Ballestrero:2008gf}
A.~Ballestrero, G.~Bevilacqua and E.~Maina, \emph{{A Complete parton level
  analysis of boson-boson scattering and ElectroWeak Symmetry Breaking in
  $\ell\nu$ + four jets production at the LHC}},
  \href{https://doi.org/10.1088/1126-6708/2009/05/015}{\emph{JHEP} {\bfseries
  05} (2009) 015} [\href{https://arxiv.org/abs/0812.5084}{{\ttfamily
  0812.5084}}].

\bibitem{Denner:2024xul}
A.~Denner, D.~Lombardi and C.~Schwan, \emph{{Double-pole approximation for
  leading-order semi-leptonic vector-boson scattering at the LHC}},
  \href{https://doi.org/10.1007/JHEP08(2024)146}{\emph{JHEP} {\bfseries 08}
  (2024) 146} [\href{https://arxiv.org/abs/2406.12301}{{\ttfamily
  2406.12301}}].

\bibitem{Denner:2005fg}
A.~Denner, S.~Dittmaier, M.~Roth and L.H.~Wieders, \emph{{Electroweak
  corrections to charged-current ${e}^+ {e}^- \to$ 4 fermion processes:
  Technical details and further results}},
  \href{https://doi.org/10.1016/j.nuclphysb.2011.09.001,
  10.1016/j.nuclphysb.2005.06.033}{\emph{Nucl. Phys.} {\bfseries B724} (2005)
  247} [\href{https://arxiv.org/abs/hep-ph/0505042}{{\ttfamily
  hep-ph/0505042}}].

\bibitem{Denner:2000bj}
A.~Denner, S.~Dittmaier, M.~Roth and D.~Wackeroth, \emph{{Electroweak radiative
  corrections to ${e}^+ {e}^- \to {W W} \to$ 4 fermions in double pole
  approximation: The RACOONWW approach}},
  \href{https://doi.org/10.1016/S0550-3213(00)00511-3}{\emph{Nucl. Phys.}
  {\bfseries B587} (2000) 67}
  [\href{https://arxiv.org/abs/hep-ph/0006307}{{\ttfamily hep-ph/0006307}}].

\bibitem{Catani:1996vz}
S.~Catani and M.H.~Seymour, \emph{{A general algorithm for calculating jet
  cross-sections in NLO QCD}},
  \href{https://doi.org/10.1016/S0550-3213(96)00589-5}{\emph{Nucl. Phys.}
  {\bfseries B485} (1997) 291}
  [\href{https://arxiv.org/abs/hep-ph/9605323}{{\ttfamily hep-ph/9605323}}].

\bibitem{Dittmaier:1999mb}
S.~Dittmaier, \emph{{A general approach to photon radiation off fermions}},
  \href{https://doi.org/10.1016/S0550-3213(99)00563-5}{\emph{Nucl. Phys.}
  {\bfseries B565} (2000) 69}
  [\href{https://arxiv.org/abs/hep-ph/9904440}{{\ttfamily hep-ph/9904440}}].

\bibitem{Catani:2002hc}
S.~Catani, S.~Dittmaier, M.H.~Seymour and Z.~Tr\'ocs\'anyi, \emph{{The dipole
  formalism for next-to-leading order QCD calculations with massive partons}},
  \href{https://doi.org/10.1016/S0550-3213(02)00098-6}{\emph{Nucl. Phys.}
  {\bfseries B627} (2002) 189}
  [\href{https://arxiv.org/abs/hep-ph/0201036}{{\ttfamily hep-ph/0201036}}].

\bibitem{Dittmaier:2008md}
S.~Dittmaier, A.~Kabelschacht and T.~Kasprzik, \emph{{Polarized QED splittings
  of massive fermions and dipole subtraction for non-collinear-safe
  observables}},
  \href{https://doi.org/10.1016/j.nuclphysb.2008.03.010}{\emph{Nucl. Phys.}
  {\bfseries B800} (2008) 146}
  [\href{https://arxiv.org/abs/0802.1405}{{\ttfamily 0802.1405}}].

\bibitem{Actis:2012qn}
S.~Actis, A.~Denner, L.~Hofer, A.~Scharf and S.~Uccirati, \emph{{Recursive
  generation of one-loop amplitudes in the Standard Model}},
  \href{https://doi.org/10.1007/JHEP04(2013)037}{\emph{JHEP} {\bfseries 04}
  (2013) 037} [\href{https://arxiv.org/abs/1211.6316}{{\ttfamily 1211.6316}}].

\bibitem{Denner:2016kdg}
A.~Denner, S.~Dittmaier and L.~Hofer, \emph{{COLLIER: a fortran-based Complex
  One-Loop LIbrary in Extended Regularizations}},
  \href{https://doi.org/10.1016/j.cpc.2016.10.013}{\emph{Comput. Phys. Commun.}
  {\bfseries 212} (2017) 220}
  [\href{https://arxiv.org/abs/1604.06792}{{\ttfamily 1604.06792}}].

\bibitem{Berends:1994pv}
F.A.~Berends, R.~Pittau and R.~Kleiss, \emph{{All electroweak four fermion
  processes in electron-positron collisions}},
  \href{https://doi.org/10.1016/0550-3213(94)90297-6}{\emph{Nucl. Phys.}
  {\bfseries B424} (1994) 308}
  [\href{https://arxiv.org/abs/hep-ph/9404313}{{\ttfamily hep-ph/9404313}}].

\bibitem{Denner:1999gp}
A.~Denner, S.~Dittmaier, M.~Roth and D.~Wackeroth, \emph{{Predictions for all
  processes ${e}^+ {e}^- \to $ 4 fermions $+ \gamma$}},
  \href{https://doi.org/10.1016/S0550-3213(99)00437-X}{\emph{Nucl. Phys.}
  {\bfseries B560} (1999) 33}
  [\href{https://arxiv.org/abs/hep-ph/9904472}{{\ttfamily hep-ph/9904472}}].

\bibitem{Dittmaier:2002ap}
S.~Dittmaier and M.~Roth, \emph{{LUSIFER: A LUcid approach to six FERmion
  production}},
  \href{https://doi.org/10.1016/S0550-3213(02)00640-5}{\emph{Nucl. Phys.}
  {\bfseries B642} (2002) 307}
  [\href{https://arxiv.org/abs/hep-ph/0206070}{{\ttfamily hep-ph/0206070}}].

\bibitem{Nagy:2003tz}
Z.~Nagy, \emph{{Next-to-leading order calculation of three jet observables in
  hadron hadron collision}},
  \href{https://doi.org/10.1103/PhysRevD.68.094002}{\emph{Phys. Rev. D}
  {\bfseries 68} (2003) 094002}
  [\href{https://arxiv.org/abs/hep-ph/0307268}{{\ttfamily hep-ph/0307268}}].

\bibitem{ParticleDataGroup:2024cfk}
{\scshape Particle Data Group} collaboration, \emph{{Review of particle
  physics}}, \href{https://doi.org/10.1103/PhysRevD.110.030001}{\emph{Phys.
  Rev. D} {\bfseries 110} (2024) 030001}.

\bibitem{Heinemeyer:2013tqa}
{\scshape LHC Higgs Cross Section Working Group} collaboration, \emph{{Handbook
  of LHC Higgs Cross Sections: 3. Higgs Properties}},  (Geneva), CERN, 2013
  [\href{https://arxiv.org/abs/1307.1347}{{\ttfamily 1307.1347}}].

\bibitem{NNPDF:2024djq}
{\scshape NNPDF} collaboration, \emph{{Photons in the proton: implications for
  the LHC}}, \href{https://doi.org/10.1140/epjc/s10052-024-12731-8}{\emph{Eur.
  Phys. J. C} {\bfseries 84} (2024) 540}
  [\href{https://arxiv.org/abs/2401.08749}{{\ttfamily 2401.08749}}].

\bibitem{Andersen:2014efa}
J.R.~Andersen et~al., \emph{{Les Houches 2013: Physics at TeV Colliders:
  Standard Model Working Group Report}},  2014
  [\href{https://arxiv.org/abs/1405.1067}{{\ttfamily 1405.1067}}].

\bibitem{Buckley:2014ana}
A.~Buckley, J.~Ferrando, S.~Lloyd, K.~Nordstr{\"o}m, B.~Page, M.~R{\"u}fenacht
  et~al., \emph{{LHAPDF6: parton density access in the LHC precision era}},
  \href{https://doi.org/10.1140/epjc/s10052-015-3318-8}{\emph{Eur. Phys. J.}
  {\bfseries C75} (2015) 132}
  [\href{https://arxiv.org/abs/1412.7420}{{\ttfamily 1412.7420}}].

\bibitem{ATLAS:2018tav}
{\scshape ATLAS} collaboration, \emph{{Prospective study of vector boson
  scattering in WZ fully leptonic final state at HL-LHC}},
  ATL-PHYS-PUB-2018-023.

\bibitem{CMS:2021qzzv1}
{\scshape CMS} collaboration, \emph{{Evidence for WW/WZ vector boson scattering
  in the decay channel \ensuremath{\ell}\ensuremath{\nu}qq produced in
  association with two jets in proton-proton collisions at
  $\sqrt{s}={}$13\,TeV}},  \href{https://arxiv.org/abs/2112.05259v1}{{\ttfamily
  2112.05259v1}}.

\bibitem{Dokshitzer:1997in}
Y.L.~Dokshitzer, G.D.~Leder, S.~Moretti and B.R.~Webber, \emph{{Better jet
  clustering algorithms}},
  \href{https://doi.org/10.1088/1126-6708/1997/08/001}{\emph{JHEP} {\bfseries
  08} (1997) 001} [\href{https://arxiv.org/abs/hep-ph/9707323}{{\ttfamily
  hep-ph/9707323}}].

\bibitem{Wobisch:1998wt}
M.~Wobisch and T.~Wengler, \emph{{Hadronization corrections to jet
  cross-sections in deep inelastic scattering}},  in \emph{{Workshop on Monte
  Carlo Generators for HERA Physics (Plenary Starting Meeting)}}, pp.~270--279,
  4, 1998 [\href{https://arxiv.org/abs/hep-ph/9907280}{{\ttfamily
  hep-ph/9907280}}].

\bibitem{Cacciari:2008gp}
M.~Cacciari, G.P.~Salam and G.~Soyez, \emph{{The anti-$k_t$ jet clustering
  algorithm}}, \href{https://doi.org/10.1088/1126-6708/2008/04/063}{\emph{JHEP}
  {\bfseries 04} (2008) 063} [\href{https://arxiv.org/abs/0802.1189}{{\ttfamily
  0802.1189}}].

\bibitem{Denner:2000jv}
A.~Denner and S.~Pozzorini, \emph{{One loop leading logarithms in electroweak
  radiative corrections. 1.~Results}},
  \href{https://doi.org/10.1007/s100520100551}{\emph{Eur. Phys. J.} {\bfseries
  C18} (2001) 461} [\href{https://arxiv.org/abs/hep-ph/0010201}{{\ttfamily
  hep-ph/0010201}}].

\bibitem{Accomando:2006hq}
E.~Accomando, A.~Denner and S.~Pozzorini, \emph{{Logarithmic electroweak
  corrections to $e^+ e^-\to\nu_e \bar{\nu}_e W^+ W^-$}},
  \href{https://doi.org/10.1088/1126-6708/2007/03/078}{\emph{JHEP} {\bfseries
  03} (2007) 078} [\href{https://arxiv.org/abs/hep-ph/0611289}{{\ttfamily
  hep-ph/0611289}}].

\bibitem{Pagani:2021vyk}
D.~Pagani and M.~Zaro, \emph{{One-loop electroweak Sudakov logarithms: a
  revisitation and automation}},
  \href{https://doi.org/10.1007/JHEP02(2022)161}{\emph{JHEP} {\bfseries 02}
  (2022) 161} [\href{https://arxiv.org/abs/2110.03714}{{\ttfamily
  2110.03714}}].

\bibitem{Lindert:2023fcu}
J.M.~Lindert and L.~Mai, \emph{{Logarithmic EW corrections at one-loop}},
  \href{https://doi.org/10.1140/epjc/s10052-024-13430-0}{\emph{Eur. Phys. J. C}
  {\bfseries 84} (2024) 1084}
  [\href{https://arxiv.org/abs/2312.07927}{{\ttfamily 2312.07927}}].

\bibitem{Denner:2024yut}
A.~Denner and S.~Rode, \emph{{Automated resummation of electroweak Sudakov
  logarithms in diboson production at future colliders}},
  \href{https://doi.org/10.1140/epjc/s10052-024-12879-3}{\emph{Eur. Phys. J. C}
  {\bfseries 84} (2024) 542}
  [\href{https://arxiv.org/abs/2402.10503}{{\ttfamily 2402.10503}}].

\end{thebibliography}\endgroup

\newpage
\newpage
\appendix

\end{document}